\documentclass[transmag]{IEEEtran}
\usepackage{cite}
\usepackage{latexsym}
\usepackage{graphicx}
\usepackage{amsfonts,amssymb,amsmath}
\usepackage{algorithmic}
\usepackage{array}
\usepackage{stfloats}
\usepackage{url}
\usepackage{lipsum}
\usepackage{upgreek}
\usepackage{hhline}
\usepackage{enumerate}
\usepackage{mathtools}
\usepackage{ulem}

\DeclareMathOperator*{\argminA}{arg\,min} % Jan Hlavacek

\def\BibTeX{{\rm B\kern-.05em{\sc i\kern-.025em b}\kern-.08em T\kern-.1667em\lower.7ex\hbox{E}\kern-.125emX}}
\begin{document}

\title{Closed-Form Solution for Scaling a Wireless Acoustic Sensor Network}

% \title{Close-Form Solution for Calibrating an Acoustic Sensor}

\author{Kashyap~Patel, Anton Kovalyov, and~Issa~Panahi~\IEEEmembership{}
\thanks{Patent on this work is pending. The authors are with the Department
of Electrical and Computer Engineering, University of Texas at Dallas, Richardson, TX-75080, USA. \newline
Corresponding authors: \newline
Kashyap Patel: patelkashyap@utdallas.edu; \newline 
Anton Kovalyov: anton.kovalyov@utdallas.edu;\newline
Issa Panahi: imp015000@utdallas.edu
}}%

\IEEEtitleabstractindextext{\begin{abstract} 
This study presents a closed-form solution for localizing and synchronizing an acoustic sensor node with respect to a Wireless Acoustic Sensor Network (WASN). The aim is to allow efficient scaling of a WASN by individually calibrating newly joined sensor nodes instead of recalibrating the entire array. A key contribution is that the sensor to be calibrated does not need to include a built-in emitter. The proposed method uses signals emitted from spatially distributed sources to compute time difference of arrival (TDOA) measurements between the existing WASN and a new sensor. The problem is then modeled as a set of multivariate nonlinear TDOA equations. Through a simple transformation, the nonlinear TDOA equations are converted into a system of linear equations. Then, weighted least squares (WLS) is applied to find an accurate estimate of the calibration parameters. Signal sources can either be known emitters within the existing WASN or arbitrary sources in the environment, thus allowing for flexible applicability in both active and passive calibration scenarios. Simulation results under various conditions show high joint localization and synchronization performance, often comparable to the Cramér-Rao lower bound (CRLB).

\end{abstract}
\begin{IEEEkeywords}
Calibration, wireless acoustic sensor networks, localization, synchronization, TDOA, weighted least squares.
\end{IEEEkeywords}}

\maketitle

\section{INTRODUCTION}

\IEEEPARstart{W}{ireless} acoustic sensor networks (WASNs) can be deployed to determine the spatio-temporal composition of an acoustic field. A WASN is here described as an ad-hoc array of spatially distributed acoustic sensor nodes interconnected by a wireless medium, with each node including a processor, a wireless transmitter and receiver, a sound sensor, and, optionally, an acoustic emitter. WASNs are used in source localization \cite{emitterLoc_All_methods}, target tracking \cite{sensor_networks}, and beamforming applications \cite{beamfroming}. These applications generally require accurate estimation of sensor locations, i.e., calibration, as well as estimation of sensor clock offsets with respect to some central reference of time, i.e., synchronization\footnote{Clock drift correction is also a necessary component of systems involving clock synchronization between nodes. Well-known solutions include the Network Time Protocol (NTP) and the Global Positioning System (GPS). Other solutions specific to WASNs can be found in \cite{drift1, drift2}}. In this work, we are particularly interested in deriving a closed-form solution for joint localization and synchronization of a sensor node, which does not include a built-in emitter, with respect to a previously calibrated WASN. The objective is to both provide an efficient means to scale a WASN as new sensors join the network and a means to add individual sensors which may not include a built-in emitter.\par

Related work includes \cite{haddad2015robust, survey, barcelo2019self, beepbeep, simul_ranging, distributed, self_loc, kovalyov2021joint, wang2019doa, diffuse, energy, passive, plinge2017passive, jacob2013doa}. In \cite{haddad2015robust}, Haddad et al. propose a robust time-of-arrival (TOA) estimation algorithm along with a least-square (LS) error minimization technique for localizing an acoustic sensor in a reverberant environment. However, their method requires the emission of controlled signals from synchronized actuators placed at known strategic locations, which limits its applicability to only a small subset of practical scenarios. In the early stages of research, the problem of acoustic sensor localization was interchangeably referred to as WASN calibration \cite{survey, barcelo2019self}. A lot of literature exists on the topic of WASN calibration. One can divide the proposed methods into two categories: \textit{active} and \textit{passive}.\par

In the more popular active methods, individual WASN nodes include an emitter device that generates a dedicated signal whenever network calibration is performed. Related work includes \cite{beepbeep, simul_ranging, distributed, self_loc, kovalyov2021joint, wang2019doa}. The ``BeepBeep" system, proposed by Peng et al. in \cite{beepbeep}, provides a method to estimate the range between two asynchronous devices. Each device includes a sensor and an emitter. The devices emit a special ``Beep" signal sequentially; then, the sensors' time of arrival (TOA) measurements are used to estimate the range between the devices. The BeepBeep system was later extended by Cobos et al. \cite{simul_ranging} to allow simultaneous emission of Beep signals between network nodes, followed by converting range estimates to absolute node positions through multidimensional scaling (MDS). Raykar et al. \cite{distributed} proposed a method that further refines node localization by estimating sensor and emitter positions within each node. Pertila et al. \cite{self_loc} considered each node a device with multiple synchronized sensors and one emitter. Their calibration procedure is similar to that of BeepBeep. The difference is that additional knowledge of sensor network geometry within each node is applied to improve localization results of individual sensors and emitters. Wang et al. \cite{wang2019doa} considered separate emitter nodes at known locations and sensor array nodes at unknown locations. Their method would estimate Direction of Arrival (DOA) measurements between emitter and sensor array nodes followed by applying Artificial Bee Colony (ABC) optimization to find the locations of the sensors. Recently, Kovalyov et al. \cite{kovalyov2021joint} proposed a method for joint localization and synchronization of two arrays of sensors and emitters by gathering TOA measurements between each sensor and emitter followed by applying particle swarm optimization (PSO) to find orientation, translation and synchronization parameters of one array with respect to the other. Active calibration methods generally attain high performance. However, the existence of dedicated emitters in each node increases the cost of the system. Furthermore, the emission of calibration signals may be too disruptive in certain applications.\par 

% WANG Method
% Wang et al. derived theoretical DOAs of sound sources relative to nodes based on 3D rotation matrices and translation vectors and estimated corresponding measured DOAs using TDOA. They formulated node geometry calibration as a non-convex minimization problem to minimize the mismatch between theoretical and measured DOAs, which was solved using the ABC algorithm. The approach assumes known intra-node geometry and positions of a few sources, with each node having multiple microphones.

% IN short
% Wang et al. \cite{wang2019doa} presented a method to localize the nodes using an iterative optimization approach of ABC on theoretical and estimated DOA measurements, knowing the positions of a few sound sources and intra-node geometry, each consisting of several sensors.

% Hadada et al. in \cite{haddad2015robust} presented related work in a single acoustic sensor localization method. This approach proposes a robust time-of-arrival (TOA) estimation algorithm and employs a least-square (LS) error minimization technique for acoustic sensor localization in the reverberant environment. However, this method requires the emission of controlled signals using synchronized acoustic sources placed at known strategic locations, making it difficult or unrealistic to implement in some cases.

Passive methods, on the contrary, use arbitrary signals in the environment to calibrate a WASN. Related work includes \cite{diffuse, energy, passive, plinge2017passive, jacob2013doa}. McCowan et al. \cite{diffuse} proposed a method for localizing synchronized sensors in a diffuse noise environment. Their method first estimates pairwise sensor ranges by fitting measured noise coherence with its theoretical model, followed by MDS to find the absolute position of sensors. On the other hand, the work in \cite{energy, passive} uses signals emitted from arbitrary sources. Chen et al. \cite{energy} estimated the position of sensors and arbitrary emitters simultaneously from signal energy measurements. Wozniak and Kowalczyk \cite{passive} proposed a two-step calibration method for distributed systems where each node is a sensor array by itself. Their method first estimates the geometry of the distributed system using DOA measurements observed individually at each node. This estimate and time difference of arrival (TDOA) measurements between nodes are combined to localize individual sensors and involved emitters and find the synchronization offsets between the nodes. It should be noted that although passive calibration methods are preferred from a practical point of view, their performance is generally less robust when compared to active calibration methodology.\par 

Applications involving WASNs generally require a large number of calibrated sensors to achieve good performance. In practice, not all sensors might be available at the start of the application. Moreover, in certain scenarios, sensor nodes may come and leave at any time, or their locations may change. Let us consider a properly calibrated WASN. Adding a new sensor would require estimating its position and synchronization offset with respect to the WASN. One solution is to apply one of the previously described methods to recalibrate the entire network, including the new sensor. However, in case of a new sensor entering the WASN, recalibration of the entire network may entail not only computational burden but also introduce new errors since the original WASN is assumed to have already been properly calibrated. In such situations, it is preferable to estimate the calibration and synchronization parameters of the new sensor only, thus not altering previous knowledge of network geometry and synchronization offsets between nodes.\par

\begin{figure*}[!t]
\begin{minipage}[b]{.48\linewidth}
  \centering
  \centerline{\includegraphics[width=8cm]{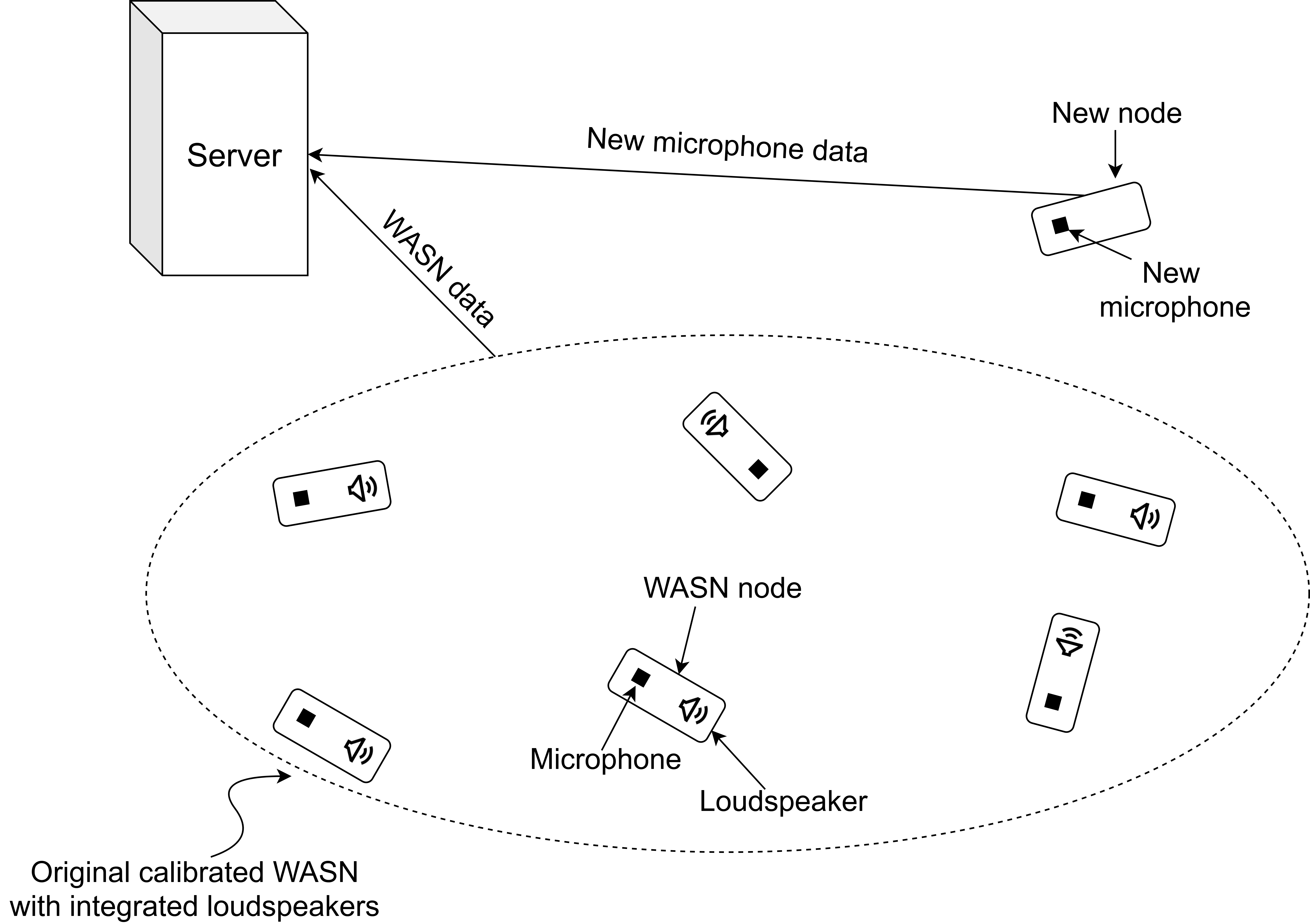}}
  \centerline{\small (a) Active calibration}\medskip
  \label{fig:DSA(a)}
\end{minipage}
\hfill
\begin{minipage}[b]{0.48\linewidth}
  \centering
  \centerline{\includegraphics[width=8cm]{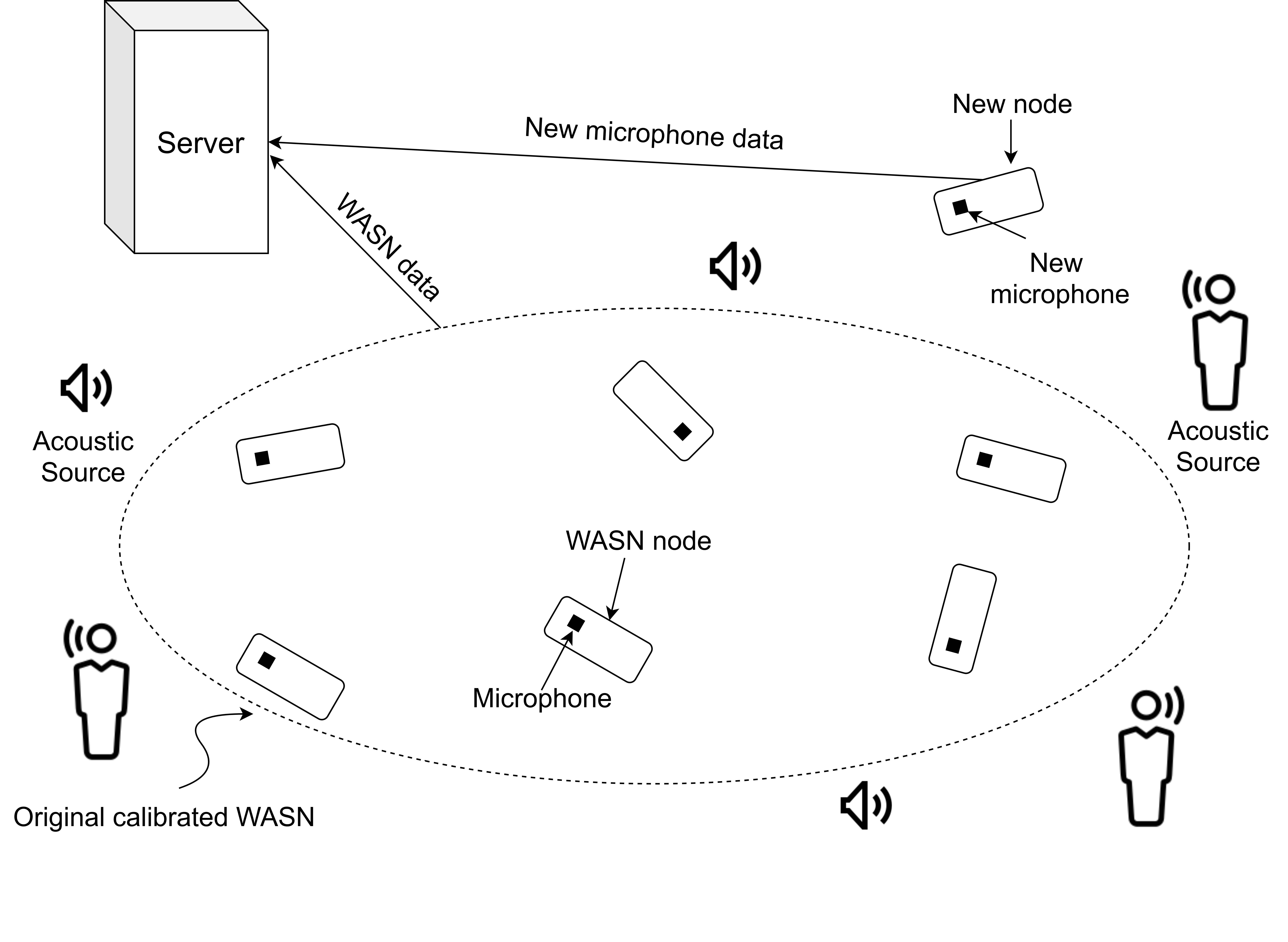}}
  \centerline{\small (b) Passive calibration}\medskip
  \label{fig:DSA(b)}
\end{minipage}
\caption{Scaling a WASN by jointly localizing and synchronizing newly joined acoustic sensor nodes. (a) Active calibration: built-in emitters within calibrated WASN nodes play dedicated calibration signals. (b) Passive calibration: takes advantage of signals produced by arbitrary sources in the surrounding environment to calibrate sensors, i.e., people speaking.}
\label{fig:DSA}
\end{figure*}

To address the concerns discussed above, we propose an efficient closed-form solution for scaling a WASN through individual joint calibration and synchronization of newly added sensor nodes to the network. The proposed method uses acoustic signals emitted from spatially distributed sources to gather TDOA measurements between the existing WASN and the new sensor, where TDOAs can be found by employing any conventional TDOA estimation method as in \cite{GCC, TDOA_multiple, TDOA_DL}. These TDOA measurements, combined with knowledge of network geometry and emitter positions, are used to estimate the location and synchronization offset of the new sensor with respect to the WASN. This estimation problem is modeled as finding the closed-form solution to a set of multivariate nonlinear TDOA equations. Through a simple transformation, the nonlinear TDOA equations are converted into a system of linear equations. Then, WLS is applied to estimate a solution that is robust to errors in all type of measurements, that is, TDOA estimates, sensor position estimates and emitter position estimates. Derivation of the Cramér-Rao lower bound (CRLB) for analysis is also presented in this paper. When evaluated under various simulation conditions, results show that the proposed method achieves high joint localization and synchronization performance, often comparable to the CRLB.\par

It should be noted that knowledge of network geometry is implied since we assume that the original WASN is properly calibrated. Knowledge of emitter positions, on the other hand, can come in two different ways: (1) emitter positions are known a priori, i.e., they are built-in devices within calibrated sensor nodes (see Fig.\ref{fig:DSA} (a)); (2) emitters are arbitrary sources in the environment (see Fig. \ref{fig:DSA} (b)) whose positions were estimated\footnote{In practice, to avoid problems related to source mismatch, the same signal used to localize the source should be used to measure TDOA between the network and the new sensor.} by the original WASN applying passive source localization methodology \cite{source_srp, source_tdoa_simple, source_tdoa_error}. Consequently, the proposed method can be employed in either active or passive calibration scenarios.\par

Let us consider a real-world application of the proposed method, e.g., a smart conference room with a built-in WASN that is assumed to be correctly calibrated. The WASN facilitates multi-channel signal processing applications, such as noise and interference suppression; speech enhancement; source positioning, identification, tracking, and targeting for audio/video conference events. In this scenario, available personal digital assistants (PDAs) such as smartphones, laptops, tablets, or smartwatches equipped with microphones and wireless transmitters can be integrated into the existing WASN by transmitting their audio input signals to the WASN server, with the objective of enhancing the performance of the aforementioned applications. The proposed method could then be used to calibrate and synchronize the PDA microphone sensors with the WASN via an active or a passive calibration approach, as illustrated by the two examples in Fig. \ref{fig:DSA}.\par

This paper is structured as follows. The problem of sensor calibration is described in detail in Section \ref{sec:formulation}. A closed-form solution based on WLS, as well as derivation of the CRLB, are presented in Section \ref{sec:cf}. Simulation results are reported in Section \ref{sec:results}. Finally, Section \ref{sec:conclusion} concludes the paper.\par

By convention, vectors in this paper are column vectors. They are denoted by lowercase bold letters/symbols. Matrices are denoted by upper case bold letters/symbols. $\mathbf{A}[i,j]$ denotes an element corresponding to the $i^{th}$ row and $j^{th}$ column in matrix $\mathbf{A}$. $[\, \cdot\, ]_{N \times M}$ defines a vector/matrix of size $N \times M$. $\mathbf{I}_N$ is the identity matrix of size $N \times N$. $\mathbf{0}$ is a vector/matrix of zeros. $(\cdot)^{T}$ denotes transposition. $(\cdot)^{-1}$ denotes inverse. $\mathbb{E}[\, \cdot\, ]$ denotes expectation. $||\mathbf{x}||$ is the Euclidean norm of $\mathbf{x}$. $\text{tr}(\mathbf{X})$ is the trace of $\mathbf{X}$. $\text{blkdiag} \left\{ [\, \cdot\, ], \ldots, [\, \cdot\, ] \right\}$ is a block diagonal matrix formed from a list of vectors/matrices. $\tilde{(\cdot)}$ denotes a known estimate. The accent $\hat{(\cdot)}$ denotes an unknown estimate that needs to be found. $\Delta(\cdot)$ denotes additive noise modeled as a zero-mean random variable. $\mathcal{N}(\boldsymbol{\mu}, \boldsymbol{\Sigma})$ denotes normal distribution with mean $\boldsymbol{\mu}$ and covariance $\boldsymbol{\Sigma}$. Finally, $\mathcal{U}(a,b)$ denotes uniform distribution whith values in the interval $[a,b]$.

\section{Problem Formulation} \label{sec:formulation}
Let $\mathbf{s}_1, ..., \mathbf{s}_M$ be the $D$-dimensional vector positions of $M$ sensors belonging to a calibrated WASN. Similarly, let $\mathbf{u}_1, ..., \mathbf{u}_N$ be the $D$-dimensional vector positions of $N$ emitters. Known estimates of sensor and emitter positions are modeled by
\begin{equation} \label{eq:si}
    \mathbf{\tilde{s}}_i = \mathbf{s}_i + \Delta \mathbf{s}_i\;, \quad i = 1,\ldots,M
\end{equation}
and
\begin{equation} \label{eq:uj}
    \mathbf{\tilde{u}}_j = \mathbf{u}_j + \Delta \mathbf{u}_j\;, \quad j = 1,\ldots,N \;.
\end{equation}
respectively. Consider a new sensor being added to the WASN. Let $\mathbf{p}$ be the $D$-dimensional position of the new sensor. Let $\tau_p$ be the synchronization offset between the new sensor and the calibrated WASN. Since the new sensor and the original WASN are not synchronized, the noise-free TDOA of the signal originating at $\mathbf{u}_j$ when received at the sensors corresponding to $\mathbf{s}_i$ and $\mathbf{p}$ is given by
\begin{equation}\label{eq:tau_ij}
    \tau_{ij} = \dfrac{||\mathbf{u}_j - \mathbf{s}_i|| - ||\mathbf{u}_j - \mathbf{p}||}{c} + \tau_p \;,
\end{equation}

% Note that the parametric formulation of the TDOA equation in (\ref{eq:tau_ij}) for acoustic sensor localization and synchronization differs from that of source localization \cite{source_tdoa_error}. In source localization, sensors are assumed to be synchronized and the source position is determined from the intersection of hyperbolas with sensor positions as foci On the other hand, (\ref{eq:tau_ij}) in the context of sensor localization and synchronization does not involve hyperbolic geometry and exhibits a non-linear relationship between the parameters $\mathbf{p}$ and $\tau_p$.

where $c$ is the, assumed to be known, the propagation speed of the signal. Note that when compared to the TDOA modeling in source localization \cite{source_tdoa_simple, source_tdoa_error}, where sensors are assumed to be synchronized, and the parameters of interest referring to the source position are determined from the intersection of hyperbolas with two different sensor locations as foci, the parametrization of (\ref{eq:tau_ij}) differs in that the synchronization assumption and hyperbolic geometry with respect to the parameters of interest, which are here the non-calibrated sensor position and synchronization offset, no longer apply. For mathematical convenience, let us multiply both sides of (\ref{eq:tau_ij}) by $c$ to get 
\begin{equation} \label{eq:rij0}
        r_{ij} = ||\mathbf{u}_j - \mathbf{s}_i|| - ||\mathbf{u}_j - \mathbf{p}|| + r_p \;,
\end{equation}
where $r_{ij} = c \tau_{ij}$ and $r_p = c \tau_p$. In practice, estimates of (\ref{eq:rij0}) may be corrupted due to many reasons, e.g., synchronization jitter between the internal clocks of calibrated sensors and the global reference clock; low sampling resolution; and reverberation. Throughout this work, known estimates of (\ref{eq:rij0}) are modeled by
\begin{equation} \label{eq:rij}
    \tilde{r}_{ij} = r_{ij} + \Delta r_{ij} \;.
\end{equation}
Consequently, given knowledge of the noisy terms in (\ref{eq:si}), (\ref{eq:uj}) and (\ref{eq:rij}), as well as the second-order statistics of the additive zero-mean noise variables $\Delta \mathbf{s}_i$, $\Delta \mathbf{u}_j$ and $\Delta r_{ij}$, the aim is to derive a closed form solution that accurately calibrates the new sensor with the original WASN, that is, finding good estimates of $\mathbf{p}$ and $\tau_p$.\par

\section{Closed Form Solution} \label{sec:cf}

The system of equations in (\ref{eq:rij0}) is nonlinear in $\mathbf{p}$. This makes it difficult to solve directly. Consequently, we will apply a set of transformations to simplify the problem. Rearranging the equations in (\ref{eq:rij0}) and squaring both sides we get
\begin{equation}
    (r_{ij} - || \mathbf{u}_j - \mathbf{s}_i|| - r_p)^2 = || \mathbf{u}_j - \mathbf{p}||^2 \;.
\end{equation}
Letting
\begin{equation}
    \beta_{ij} = r_{ij} - ||\mathbf{u}_j - \mathbf{s}_i|| \;,
\end{equation}
followed by expanding the square on both sides, results in the following set of equations
\begin{equation} \label{eq:non_linear}
    \beta_{ij}^{2} - 2\beta_{ij} r_p + r_p^{ 2} = \mathbf{u}_j^{T} \mathbf{u}_j - 2\mathbf{u}_j^{T} \mathbf{p} + \mathbf{p}^{T} \mathbf{p} \;.
\end{equation}
Then, for $i = 1,...,M$, subtracting the term for $j = N$ from (\ref{eq:non_linear}) yields
\begin{equation} \label{eq:linear}
    \begin{multlined}
        \frac{1}{2} \left[ \beta_{ij}^{2} - \beta_{iN}^{2} - \mathbf{u}_j^{T} \mathbf{u}_j + \mathbf{u}_N^T \mathbf{u}_N \right] \\ = 
     (-\mathbf{u}_j + \mathbf{u}_N)^T\mathbf{p} + (\beta_{ij} - \beta_{iN}) r_p \;, \\ 
    j = 1, \ldots, N-1 \;.
    \end{multlined}
\end{equation}
This is now a system of linear equations in the unknown variables, and it follows that the constraints $M \geq 1$ and $N \geq D + 2$ must be satisfied for estimation to be possible.  When $M = 1$ and $N = D + 2$, we have a linear system of $D + 1$ equations and $D + 1$ unknowns, which can be solved directly. But results will not be satisfactory, even under low noise conditions. Allowing redundant measurements instead can significantly improve estimation performance. Consequently, we propose applying WLS to solve the problem, given redundant measurements.\par

Since only noisy measurements are available, (\ref{eq:linear}) should be rewritten to construct an error vector. For mathematical convenience, let us first define
\begin{equation} \label{eq:bij_measurement}
    \tilde{\beta}_{ij} = \tilde{r}_{ij} - ||\mathbf{\tilde{u}}_j - \mathbf{\tilde{s}}_i|| \;.
\end{equation}
Note that $\tilde{\beta}_{ij}$ is itself a noisy measurement that can also be expressed as
\begin{equation} \label{eq:bij}
    \begin{aligned}
        \tilde{\beta}_{ij} &= \beta_{ij} + \Delta \beta_{ij} \\
                   &= \beta_{ij} + \Delta r_{ij} - \Delta \eta_{ij} \;,
    \end{aligned}
\end{equation}
where $\Delta \eta_{ij}$ refers to the additive noise in
\begin{equation}\label{eq:uj_si}
    ||\mathbf{\tilde{u}}_j - \mathbf{\tilde{s}}_i|| = ||\mathbf{u}_j - \mathbf{s}_i|| + \Delta \eta_{ij} \;.
\end{equation}
An approximation of $\Delta \eta_{ij}$ is derived in Appendix \ref{appendix2}. It is given by
\begin{equation} \label{eq:delta_eta_ij}
        \Delta \eta_{ij} \approx \dfrac{(\mathbf{u}_j - \mathbf{s}_i)^T(\boldsymbol{\Delta}\mathbf{u}_j - \boldsymbol{\Delta}\mathbf{s}_i)}
        {||\mathbf{u}_j - \mathbf{s}_i||} \;.
\end{equation}
Then, upon replacing $\mathbf{u}_j$ and $\beta_{ij}$ in (\ref{eq:linear}) with their noisy counterparts, an error vector can be constructed as
\begin{equation} \label{eq:error}
    \boldsymbol{\uppsi} = \mathbf{\tilde{h}} - \mathbf{\tilde{G}}\boldsymbol{\upgamma} \;,
\end{equation}
where
\begin{align}
    \begin{split}
        \mathbf{\tilde{h}} &= \dfrac{1}{2} \label{eq:h}
        \begin{bmatrix}
            \mathbf{\tilde{h}}_1^T & \cdots & \mathbf{\tilde{h}}_M^T
        \end{bmatrix}^T \\[5pt]
        \mathbf{\tilde{h}}_i &= 
        \begin{bmatrix}
            \tilde{\beta}_{i1}^2 - \tilde{\beta}_{iN}^2 - \mathbf{\tilde{u}}_1^T \mathbf{\tilde{u}}_1 + \mathbf{\tilde{u}}_N^T \mathbf{\tilde{u}}_N \\
            \vdots \\
            \tilde{\beta}_{iN-1}^2 - \tilde{\beta}_{iN}^2 - \mathbf{\tilde{u}}_{N-1}^T \mathbf{\tilde{u}}_{N-1}^T + \mathbf{\tilde{u}}_N^T \mathbf{\tilde{u}}_N
        \end{bmatrix}\;,
    \end{split} \\[15pt]
    \begin{split}
        \mathbf{\tilde{G}} &= \label{eq:G}
        \begin{bmatrix}
            \mathbf{\tilde{G}}_1^T & \cdots & \mathbf{\tilde{G}}_M^T
        \end{bmatrix}^T\\[5pt]
        \mathbf{\tilde{G}}_i &= 
        \begin{bmatrix}
            (-\mathbf{\tilde{u}}_1 + \mathbf{\tilde{u}}_N)^T      &   \tilde{\beta}_{i1} - \tilde{\beta}_{iN}     \\
            \vdots                               &   \vdots                \\
            (-\mathbf{\tilde{u}}_{N-1} + \mathbf{\tilde{u}}_N)^T  &   \tilde{\beta}_{iN-1} - \tilde{\beta}_{iN}
        \end{bmatrix}    
    \end{split}
\end{align}
and $\boldsymbol\upgamma$ groups the unknown calibration parameters as
\begin{equation}
    \boldsymbol{\upgamma} = 
    \begin{bmatrix}
        \mathbf{p}^{ T} & r_p
    \end{bmatrix}^T \;.
\end{equation}
Under the assumption of zero-mean Gaussian noise, applying WLS, estimation of $\boldsymbol\upgamma$ becomes a minimization problem given by
\begin{equation} \label{eq:min}
    \boldsymbol{\hat{\upgamma}} = \argminA_{\boldsymbol\upgamma}\{(\mathbf{\tilde{h}} - \mathbf{\tilde{G}}\boldsymbol\upgamma)^T
                        \boldsymbol\Uppsi^{-1}(\mathbf{\tilde{h}} - \mathbf{\tilde{G}}\boldsymbol\upgamma)\} \;,
\end{equation}
where $\boldsymbol\Uppsi = \mathbb{E}[\boldsymbol\uppsi\boldsymbol\uppsi^T]$. Then, taking the derivative of the right-hand side of (\ref{eq:min}) with respect to $\boldsymbol\upgamma$ and setting the result to zero yields a linear estimator given by
\begin{equation} \label{eq:wls}
    \boldsymbol{\hat\upgamma} = (\mathbf{\tilde{G}}^T\boldsymbol{\Uppsi}^{-1}\mathbf{\tilde{G}})^{-1}\mathbf{\tilde{G}}^T\boldsymbol{\Uppsi}^{-1}\mathbf{\tilde{h}} \;.
\end{equation}
Assuming knowledge of second-order noise statistics, finding $\boldsymbol\Uppsi$ is straightforward once $\boldsymbol\uppsi$ has been restructured as a weighted sum of noise terms, which can be modeled by
\begin{equation} \label{eq:weighted_error}
    \boldsymbol{\uppsi} = \mathbf{W}_r\boldsymbol{\Delta} \mathbf{r} + 
                        \mathbf{W}_u\boldsymbol{\Delta} \mathbf{u} + 
                        \mathbf{W}_s\boldsymbol{\Delta} \mathbf{s} +
                        \mathbf{W}_\kappa\boldsymbol{\Delta\kappa} \;,
\end{equation}
where $\boldsymbol\Delta \mathbf{r}$, $\boldsymbol\Delta \mathbf{u}$, $\boldsymbol\Delta \mathbf{s}$ are vectors grouping linear noise terms as
\begin{align}
    \begin{split}
        \boldsymbol\Delta \mathbf{r} =& 
        \begin{bmatrix}
            \boldsymbol\Delta \mathbf{r}_1^T & \cdots & \boldsymbol\Delta \mathbf{r}_M^T
        \end{bmatrix}^T \\
        \boldsymbol\Delta \mathbf{r}_i =&
        \begin{bmatrix}
            \Delta r_{i1} & \cdots & \Delta r_{iN}
        \end{bmatrix}^T\;,        
    \end{split}
\end{align}
\begin{equation}
 \boldsymbol\Delta \mathbf{u} = 
     \begin{bmatrix}
             \boldsymbol{\Delta}\mathbf{u}_1^T & \cdots & \boldsymbol{\Delta}\mathbf{u}_N^T
        \end{bmatrix}^T\;, 
\end{equation}
\begin{equation}
   \boldsymbol\Delta \mathbf{s} = 
     \begin{bmatrix}
             \boldsymbol\Delta \mathbf{s}_1^T & \cdots & \boldsymbol\Delta \mathbf{s}_M^T
        \end{bmatrix}^T\;, 
\end{equation}
$\boldsymbol{\Delta \kappa}$ is a vector grouping second order noise terms and $\mathbf{W}_r$, $\mathbf{W}_u$, $\mathbf{W}_s$ and $\mathbf{W}_\kappa$ are noise weight matrices.\par
The undefined variables in (\ref{eq:weighted_error}) can be found upon substitution of the noisy measurements in (\ref{eq:error}) with their corresponding right-hand side equivalents in (\ref{eq:uj}) and (\ref{eq:bij}). Then, applying the approximation in (\ref{eq:delta_eta_ij}) and simplifying, we get 
\begin{align}
    \begin{split}
        \boldsymbol{\Delta\kappa} &=
        \begin{bmatrix}
            \boldsymbol{\Delta \kappa}_1^T &
            \cdots &
            \boldsymbol{\Delta \kappa}_M^T
        \end{bmatrix}^T  \\[5pt]
        \boldsymbol{\Delta \kappa}_i &=
        \begin{bmatrix}
            \Delta \beta_{i1}^2 - \boldsymbol\Delta \mathbf{u}_1^T \boldsymbol\Delta \mathbf{u}_1 \\
            \vdots \\
            \Delta \beta_{iN}^2 - \boldsymbol\Delta \mathbf{u}_N^T \boldsymbol\Delta \mathbf{u}_N
        \end{bmatrix} \;,
    \end{split}
\end{align}

\begin{equation} \label{eq:W_r}
    \begin{split}
        \mathbf{W}_r &= \text{blkdiag}\{{\mathbf{A}_{1}, \ldots, \mathbf{A}_{M}}\} \\[5pt]
        \mathbf{A}_{i} &= 
        \begin{bmatrix}
            a_{1}  &          &              &   -a_{N} \\
                   &   \ddots  &         &   \vdots \\
                   &     &  a_{N-1}    &   -a_{N}
        \end{bmatrix} \\[5pt]
        a_{j} &= -||\mathbf{u}_j - \mathbf{p}|| \;,
    \end{split} 
\end{equation}

\begin{equation}
    \begin{split}
        \mathbf{W}_u &=
        \begin{bmatrix}
            \mathbf{A}_{1}^T & \cdots & \mathbf{A}_{M}^T
        \end{bmatrix}^T \\[5pt] 
        \mathbf{A}_{i} &=
        \begin{bmatrix}
            \mathbf{a}_{i1}^T  &    &     &   -\mathbf{a}_{iN}^T \\
                  &   \ddots       &                    &   \vdots \\
                  &          &   \mathbf{a}_{iN-1}^T    &   -\mathbf{a}_{iN}^T
        \end{bmatrix} \\[5pt]
        \mathbf{a}_{ij} &= \dfrac{||\mathbf{u}_j - \mathbf{p}||(\mathbf{u}_j - \mathbf{s}_i)}{||\mathbf{u}_j - \mathbf{s_i}||} - (\mathbf{u}_j - \mathbf{p}) \;,
    \end{split}
\end{equation}

\begin{equation} \label{eq:W_s}
      \begin{split}
        \mathbf{W}_s &= \text{blkdiag}\{{\mathbf{A}_{1}, \ldots, \mathbf{A}_{M}}\} \\[5pt] \mathbf{A}_{i} &= 
        \begin{bmatrix}
            \mathbf{a}_{i1} - \mathbf{a}_{iN} & \cdots & \mathbf{a}_{iN-1} - \mathbf{a}_{iN}
        \end{bmatrix}^T \\[5pt]
        \mathbf{a}_{ij} &=
         -\frac{||\mathbf{u}_j - \mathbf{p}||(\mathbf{u}_j - \mathbf{s}_i)}{||\mathbf{u}_j - \mathbf{s}_i||}
    \end{split}
\end{equation}
and
\begin{equation}
      \begin{split}
        \mathbf{W}_\kappa &= \text{blkdiag}\{ \mathbf{A}_{1}, \ldots, \mathbf{A}_{M} \} \\[5pt]
        \mathbf{A}_{i} &= \dfrac{1}{2}
        \begin{bmatrix}
            1 &       &       & -1 \\
             & \ddots &  & \vdots \\
             &  &      1 & -1
        \end{bmatrix}_{N-1 \times N} .
    \end{split}
\end{equation}

Let $\mathbf{Q}_r = \mathbb{E}[\boldsymbol\Delta \mathbf{r} \boldsymbol\Delta \mathbf{r}^T]$, $\mathbf{Q}_u = \mathbb{E}[\boldsymbol\Delta \mathbf{u} \boldsymbol\Delta \mathbf{u}^T]$, $\mathbf{Q}_s = \mathbb{E}[\boldsymbol\Delta \mathbf{s} \boldsymbol\Delta \mathbf{s}^T]$ be the known noise covariance matrices. Additionally, let $\mathbf{Q}_{ru} = \mathbb{E}[\boldsymbol\Delta \mathbf{r} \boldsymbol\Delta \mathbf{u}^T]$, $\mathbf{Q}_{rs} = \mathbb{E}[\boldsymbol\Delta \mathbf{r} \boldsymbol\Delta \mathbf{s}^T]$, $\mathbf{Q}_{us} = \mathbb{E}[\boldsymbol\Delta \mathbf{u} \boldsymbol\Delta \mathbf{s}^T]$ be the known noise cross-covariance matrices. Then, neglecting the second order terms grouped by $\boldsymbol{\Delta\kappa}$ in (\ref{eq:weighted_error}), $\boldsymbol\Uppsi$ is found to be
\begin{equation}
    \begin{aligned}
        \boldsymbol{\Uppsi} = \mathbb{E}[\boldsymbol\uppsi\boldsymbol\uppsi^T] &= \boldsymbol{\Uppsi}_r + \boldsymbol{\Uppsi}_u + \boldsymbol{\Uppsi}_s + \boldsymbol{\Uppsi}_{ru} + \boldsymbol{\Uppsi}_{rs} + \boldsymbol{\Uppsi}_{us} \\[5pt]
        \boldsymbol{\Uppsi}_r &= \mathbf{W}_r\mathbf{Q}_r\mathbf{W}_r^T \\[5pt]
        \boldsymbol{\Uppsi}_u &= \mathbf{W}_u\mathbf{Q}_u\mathbf{W}_u^T \\[5pt]
        \boldsymbol{\Uppsi}_s &= \mathbf{W}_s\mathbf{Q}_s\mathbf{W}_s^T \\[5pt]
        \boldsymbol{\Uppsi}_{ru} &= \mathbf{W}_{r}\mathbf{Q}_{ru}\mathbf{W}_{u}^T + \mathbf{W}_{u}\mathbf{Q}_{ru}^T\mathbf{W}_{r}^T \\[5pt]
        \boldsymbol{\Uppsi}_{rs} &= \mathbf{W}_{r}\mathbf{Q}_{rs}\mathbf{W}_{s}^T + \mathbf{W}_{s}\mathbf{Q}_{rs}^T\mathbf{W}_{r}^T \\[5pt]
        \boldsymbol{\Uppsi}_{us} &= \mathbf{W}_{u}\mathbf{Q}_{us}\mathbf{W}_{s}^T + \mathbf{W}_{s}\mathbf{Q}_{us}^T\mathbf{W}_{u}^T \;.
    \end{aligned}
\end{equation} \par
Note that constructing $\boldsymbol\Uppsi$ requires prior knowledge of $\mathbf{s}_i$, $\mathbf{u}_j$ and $\mathbf{p}$ as shown by (\ref{eq:W_r})-(\ref{eq:W_s}). In practice, $\mathbf{s}_i$ and $\mathbf{u}_j$ can be approximated by their corresponding estimates. On the other hand, since no estimate of $\mathbf{p}$ is available, a good initial guess can be found through least squares (LS) estimation as follows
\begin{equation}\label{eq:ls}
    \boldsymbol{\hat\upgamma} =   (\mathbf{\tilde{G}}^T\mathbf{\tilde{G}})^{-1}\mathbf{\tilde{G}}^T\mathbf{\tilde{h}} \;.
\end{equation}
Then, the WLS estimator in (\ref{eq:wls}) can be applied iteratively to improve estimation further. Although, simulation results show that applying (\ref{eq:wls}) only once is usually sufficient to obtain a good final estimate of $\boldsymbol\upgamma$.

%-------------Estimator Variance ---------------

\subsection{Estimator Variance} \label{sec:estimator_variance}
The variance of the proposed WLS estimator is found following a perturbation approach. Let $\mathbf{\tilde{G}} = \mathbf{G} + \boldsymbol\Delta \mathbf{G}$, $\mathbf{\tilde{h}} = \mathbf{h} + \boldsymbol\Delta \mathbf{h}$ and $\boldsymbol{\hat\upgamma} = \boldsymbol\upgamma + \boldsymbol{\Delta \upgamma}$. Substituting into (\ref{eq:wls}) and rearranging gives
\begin{equation} \label{eq:var_perturbation}
\begin{multlined}
      (\mathbf{G} + \boldsymbol\Delta \mathbf{G})^T \boldsymbol\Uppsi^{-1} (\mathbf{G} + \boldsymbol\Delta \mathbf{G}) (\boldsymbol\upgamma + \boldsymbol\Delta\boldsymbol\upgamma) \\ = 
    (\mathbf{G} + \boldsymbol\Delta \mathbf{G})^T \boldsymbol\Uppsi^{-1} (\mathbf{h} + \boldsymbol\Delta \mathbf{h}) \;.
\end{multlined}
\end{equation}
Followed by neglecting non-linear error terms and rearranging once again, resulting in
\begin{equation} \label{eq:var_perturbation_2}
\begin{multlined}
    \mathbf{G}^{T} \boldsymbol\Uppsi^{-1} \mathbf{G} \boldsymbol\Delta \boldsymbol\upgamma \\ = \mathbf{G}^{T} \boldsymbol\Uppsi^{-1}[(\mathbf{h} - \mathbf{G} \boldsymbol\upgamma) + (\boldsymbol\Delta \mathbf{h} - \boldsymbol\Delta \mathbf{G} \boldsymbol\upgamma)] \\ +\boldsymbol\Delta \mathbf{G}^T \boldsymbol\Uppsi^{-1} (\mathbf{h} - \mathbf{G} \boldsymbol\upgamma) \;.
\end{multlined}
\end{equation}
Noting that $ \mathbf{h} = \mathbf{G} \boldsymbol\upgamma $ and $\boldsymbol\uppsi = \boldsymbol\Delta\mathbf{h} - \boldsymbol\Delta\mathbf{G}\boldsymbol\upgamma $
allows simplifying (\ref{eq:var_perturbation_2}) into
\begin{equation}
    \boldsymbol\Delta\boldsymbol\upgamma = (\mathbf{G}^{T} \boldsymbol\Uppsi^{-1} \mathbf{G})^{-1} \mathbf{G}^{T} \boldsymbol\Uppsi^{-1} \boldsymbol\uppsi \;.
\end{equation}
It follows that the covariance matrix of $\boldsymbol{\hat{\upgamma}}$ is given by
\begin{equation}\label{eq:wls-T}
    \text{cov}(\boldsymbol{\hat{\upgamma}}) = \mathbb{E}[\boldsymbol{\Delta\upgamma\Delta\upgamma}^T] = (\mathbf{G}^{T} \boldsymbol\Uppsi^{-1} \mathbf{G})^{-1} \;,
\end{equation}
where the definition $\boldsymbol{\Uppsi} = \mathbb{E}[\boldsymbol\uppsi\boldsymbol\uppsi^T]$ was used.

% --------------CRLB ----------------

\subsection{Cram\'{e}r-Rao lower bound (CRLB)}\label{sec:crlb}
The CRLB places a lower bound on the variance of an unbiased estimator \cite{estimation}. It is of interest to compare the variance of our estimator with the theoretical optimum. The CRLB is found as the inverse of the Fisher information matrix. The Fisher information matrix is given by
% In this section we derive CRLB, which places a lower bound on the variance of any unbiased estimator \cite{estimation}. Proposed WLS method uses the weights as inverse of estimated covariance matrix, hence we can assume that it is an unbiased estimator \cite{aitken1936iv}. Here, we wish to estimate the vector parameter $\boldsymbol\theta = [\boldsymbol\upgamma^T, \mathbf{u}^T, \mathbf{s}^T]^T$ based on the given measurement $\mathbf{v} = [\mathbf{r}^T, \mathbf{u}^T, \mathbf{s}^T]^T$. Note that, it is efficient to consider $\mathbf{u}$ and $\mathbf{s}$ as parameter, since it's true values are unknown. CRLB can be found as inverse of Fisher information matrix $\boldsymbol{\mathcal{J}}$, which is given by 
\begin{equation}
    \boldsymbol{\mathcal{I}}(\boldsymbol{\theta}) = -\mathbb{E}\left[ \frac{\partial^2 \ln{p(\mathbf{v}|\boldsymbol\theta)}} {\partial \boldsymbol\theta\partial\boldsymbol\theta^T} \right],
    \label{eq:fisher}
\end{equation}
where $\boldsymbol\theta$ is a vector grouping unknown parameters, $\mathbf{v}$ is a vector grouping measurements, and $p(\mathbf{v}|\boldsymbol\theta)$ is the probability density function (PDF) of $\mathbf{v}$ conditioned on $\boldsymbol\theta$.\par
In our context, the vector of measurements is
\begin{equation}
    \mathbf{v} = 
    \begin{bmatrix}
        \mathbf{\tilde{r}}^T & \mathbf{\tilde{u}}^T & \mathbf{\tilde{s}}^T
    \end{bmatrix}^T .
\end{equation}
where
\begin{equation}
    \begin{split}
        \mathbf{\tilde{r}} =& 
        \begin{bmatrix}
            \mathbf{\tilde{r}}_1^T & \cdots & \mathbf{\tilde{r}}_M^T
        \end{bmatrix}^T \\
        \mathbf{\tilde{r}}_i =&
        \begin{bmatrix}
             \tilde{r}_{i1} & \cdots &  \tilde{r}_{iN}
        \end{bmatrix}^T, 
    \end{split}
\end{equation}
\begin{equation}
    \mathbf{\tilde{u}} = 
    \begin{bmatrix}
        \mathbf{\tilde{u}}_1^T & \cdots & \mathbf{\tilde{u}}_N^T
    \end{bmatrix}^T
\end{equation}
and
\begin{equation}
    \mathbf{\tilde{s}} = 
    \begin{bmatrix}
        \mathbf{\tilde{s}}_1^T & \cdots & \mathbf{\tilde{s}}_M^T
    \end{bmatrix}^T .
\end{equation}
On the other hand, the vector of unknown parameters is
\begin{equation} \label{eq:theta}
    \boldsymbol\theta = 
    \begin{bmatrix}
        \boldsymbol{\upgamma}^T & \mathbf{u}^T & \mathbf{s}^T
    \end{bmatrix}^T \;.
\end{equation}
Note that since the true values of the emitter and sensor positions are unknown, it is necessary to include them in $\boldsymbol\theta$ of (\ref{eq:theta}) for computing an accurate lower bound.

Given the complex nature of the problem, it would be difficult to analytically derive the PDF of $\mathbf{v}$ in terms of $\boldsymbol\theta$. For simplicity, we will assume that $\mathbf{\tilde{r}}$, $\mathbf{\tilde{u}}$ and $\mathbf{\tilde{s}}$ are independent random variables in $\mathcal{N}(\mathbf{r}, \mathbf{Q}_r)$, $\mathcal{N}(\mathbf{u}, \mathbf{Q}_u)$ and $\mathcal{N}(\mathbf{s}, \mathbf{Q}_s)$, respectively. It follows that the PDF is given by
\begin{equation} \label{eq:likelihood}
    p(\mathbf{v}|\boldsymbol\theta) = p(\mathbf{\tilde{r}}|\boldsymbol\theta)p(\mathbf{\tilde{u}}|\boldsymbol\theta)p(\mathbf{\tilde{s}}|\boldsymbol\theta)\;.
\end{equation}
Applying the natural logarithm to (\ref{eq:likelihood}) we get
\begin{equation}
    \begin{aligned}
        \ln{p(\mathbf{v}|\boldsymbol\theta)} = -\frac{1}{2}[&(\mathbf{\tilde{r}} - \mathbf{r})^T \mathbf{Q}_r^{-1} (\mathbf{\tilde{r}} - \mathbf{r}) \ + \\ &(\mathbf{\tilde{u}} - \mathbf{u})^T \mathbf{Q}_u^{-1} (\mathbf{\tilde{u}} - \mathbf{u}) \ + \\
        &(\mathbf{\tilde{s}} - \mathbf{s})^T \mathbf{Q}_s^{-1} (\mathbf{\tilde{s}} - \mathbf{s})] + C \;,
    \end{aligned}
\end{equation}
where C is a constant term. Consequently, the CRLB is derived as
\begin{equation}
\begin{gathered}
     \text{CRLB} = \boldsymbol{\mathcal{I}}^{-1}(\mathbf{\theta}) =\\ \left\{-\mathbb{E}
        \begin{bmatrix}
            \dfrac{\partial^2 \ln{p(\mathbf{v}|\boldsymbol\theta)}}{\partial\boldsymbol\upgamma\partial\boldsymbol\upgamma^{T}} & 
            \dfrac{\partial^2 \ln{p(\mathbf{v}|\boldsymbol\theta)}}{\partial\boldsymbol\upgamma\partial\mathbf{u}^{T}} & 
            \dfrac{\partial^2 \ln{p(\mathbf{v}|\boldsymbol\theta)}}{\partial\boldsymbol\upgamma^{}\partial\mathbf{s}^{ T}} \\
            \dfrac{\partial^2 \ln{p(\mathbf{v}|\boldsymbol\theta)}}{\partial\mathbf{u}^{}\partial\boldsymbol\upgamma^{ T}} & 
            \dfrac{\partial^2 \ln{p(\mathbf{v}|\boldsymbol\theta)}}{\partial\mathbf{u}^{}\partial\mathbf{u}^{ T}} &
            \dfrac{\partial^2 \ln{p(\mathbf{v}|\boldsymbol\theta)}}{\partial\mathbf{u}^{}\partial\mathbf{s}^{ T}} \\
            \dfrac{\partial^2 \ln{p(\mathbf{v}|\boldsymbol\theta)}}{\partial\mathbf{s}^{}\partial\boldsymbol\upgamma^{ T}} & 
            \dfrac{\partial^2 \ln{p(\mathbf{v}|\boldsymbol\theta)}}{\partial\mathbf{s}^{}\partial\mathbf{u}^{ T}} & 
            \dfrac{\partial^2 \ln{p(\mathbf{v}|\boldsymbol\theta)}}{\partial\mathbf{s}^{}\partial\mathbf{s}^{ T}}
        \end{bmatrix}
    \right\}^{-1},
\end{gathered}
    \label{eq:crlb}
\end{equation}
where diagonal values represent the lower bound on the variance of each parameter in $\boldsymbol\theta$. More specifically, the first four diagonal values correspond to the lower bound on the variance in $\boldsymbol{\hat\upgamma}$. Solutions to the partial derivative terms in (\ref{eq:crlb}) can be found in Appendix \ref{appendix1}.

%------------Simulation results-----------
\section{Simulation Experiments} \label{sec:results}
Eight experiments, each consisting of numerous Monte Carlo (MC) simulations, were conducted to validate the performance of the proposed method under different conditions. In all experiments, we considered a WASN with $M$ sensors and $N$ emitters spanning a 3D space. The sensors and emitters were placed randomly at a distance from the origin drawn from $\mathcal{U}(0, A)$ m, where $A$ is defined as the \textit{aperture} of the WASN and emitters. The new sensor that needs to be calibrated, on the other hand, was placed randomly at a distance drawn from $\mathcal{U}(0, R)$ m, where $R$ is defined as the \textit{range} of the new sensor. The azimuth and elevation angles of all elements were drawn from $\mathcal{U}(0,2\pi)$ and $\mathcal{U}(-\pi/2,\pi/2)$ radians, respectively. All elements have been constrained to a distance of at least $5$ cm from each other. The synchronization offset between the new sensor and the WASN was drawn from $\mathcal{U}(0, 1)$ s. The speed of sound constant $c$ in (\ref{eq:tau_ij}) was set to $343$ m/s.\par

In all experiments, noisy estimates of the 3D sensor positions in (\ref{eq:si}) and 3D emitter positions in (\ref{eq:uj}) were simulated by corrupting their true coordinate values using zero-mean additive white Gaussian noise (AWGN) with standard deviations (SDs) $\sigma_s$ and $\sigma_u$, respectively. Similarly, in the first seven experiments, the TDOA estimates in (\ref{eq:rij}) were simulated by corrupting the true values in (\ref{eq:rij0}) using AWGN with SD $\sigma_r$. Consequently, we let $\mathbf{Q}_r = \sigma_r^2\mathbf{I}_{NM}$, $\mathbf{Q}_u = \sigma_u^2\mathbf{I}_{3N}$, $\mathbf{Q}_s = \sigma_s^2\mathbf{I}_{3M}$, $\mathbf{Q}_{ru} = \mathbf{0}_{NM \times 3N}$, $\mathbf{Q}_{rs} = \mathbf{0}_{NM \times 3M}$ and $\mathbf{Q}_{us} = \mathbf{0}_{3N \times 3M}$.\par 

In all experiments, we considered $N_s$ different geometric setups of the WASN, the emitters, and the new sensor. For each geometric setup, the noise was bootstrapped $N_i$ times. Two variants of the proposed WLS method, namely, WLS-1 and WLS-5, were compared with: the LS method in (\ref{eq:ls}); the estimator covariance derived in Section \ref{sec:estimator_variance}, named WLS theoretical (WLS-T); and the CRLB derived in Section \ref{sec:crlb}. The number after the hyphen in WLS-1 and WLS-5 represents the number of times (\ref{eq:wls}) is iterated. The performance of each method was measured in terms of localization and synchronization root mean square errors (RMSE). The respective localization RMSE ($\text{RMSE}_{\text{loc}}$) and synchronization RMSE ($\text{RMSE}_{\text{syn}}$) of LS, WLS-1 and WLS-5 were defined by
\begin{equation}\label{eq:RMSE_loc}
    \text{RMSE}_{\text{loc}} = \sqrt{\sum\limits_{i=1}^{N_s} \sum\limits_{j=1}^{N_i} \frac{\Vert \mathbf{p}_i - \hat{\mathbf{p}}_{ij} \Vert^2}{N_s N_i}} \; \; \text{(m)} \;
\end{equation}
and
\begin{equation}\label{eq:RMSE_syn}
    \text{RMSE}_{\text{syn}} = \sqrt{\sum\limits_{i=1}^{N_s} \sum\limits_{j=1}^{N_i} \frac{(\tau_{pi} - \hat{\tau}_{pij})^2}{N_s N_i}} \; \; \text{(s)} \;,
\end{equation}
where the subscript $i$ represents the $i$-th geometric setup and the subscript $ij$ represents the $j$-th bootstrapped noise trial of the $i$-th geometric setup. The respective performance metrics of WLS-T and CRLB were defined by  
\begin{equation}\label{eq:RMSE_loc_CRLB}
    \text{RMSE}_{\text{loc}} = \sqrt{\sum\limits_{i=1}^{N_s} \sum\limits_{d=1}^3 \frac{\text{cov}(\boldsymbol{\hat{\upgamma}}_i)[d,d]}{N_s} } \; \; (\text{m})
\end{equation}
and
\begin{equation}\label{eq:RMSE_syn_CRLB}
    \text{RMSE}_{\text{syn}} = c^{-1}\sqrt{\sum\limits_{i=1}^{N_s} \frac{\text{cov}(\boldsymbol{\hat{\upgamma}}_i)[4,4]}{N_s}} \; \; \text{(s)} \;,
\end{equation}
where $\text{cov}(\boldsymbol{\hat{\upgamma}}_i)$ is the estimate covariance of the $i$-th geometric setup, which, in the case of WLS-T, is given by (\ref{eq:wls-T}), and in the case of the CRLB, it is given by (\ref{eq:crlb}).\par

Unless otherwise specified, the parameters in all experiments were defined as follows. The number of sensors, $M$, and emitters, $N$, were set to $10$. The aperture of the WASN and emitters, $A$, and the new sensor range, $R$, were set to $1$ m. The standard deviation of the noise in sensor positions, $\sigma_s$, emitter positions, $\sigma_u$, and TDOA measurements, $\sigma_r$, were set by letting $10\log_{10}(\sigma_s) = 10\log_{10}(\sigma_u) = 10\log_{10}(\sigma_r) = -30$. The number of geometric setups, $N_s$, was set to 32. Finally, the number of times we bootstrap the noise for each geometric setup, $N_i$, was set to $10^3$.\par

\begin{figure*}
  \centering
  \begin{tabular}{ c c c c}
    \includegraphics[width=0.23\textwidth]{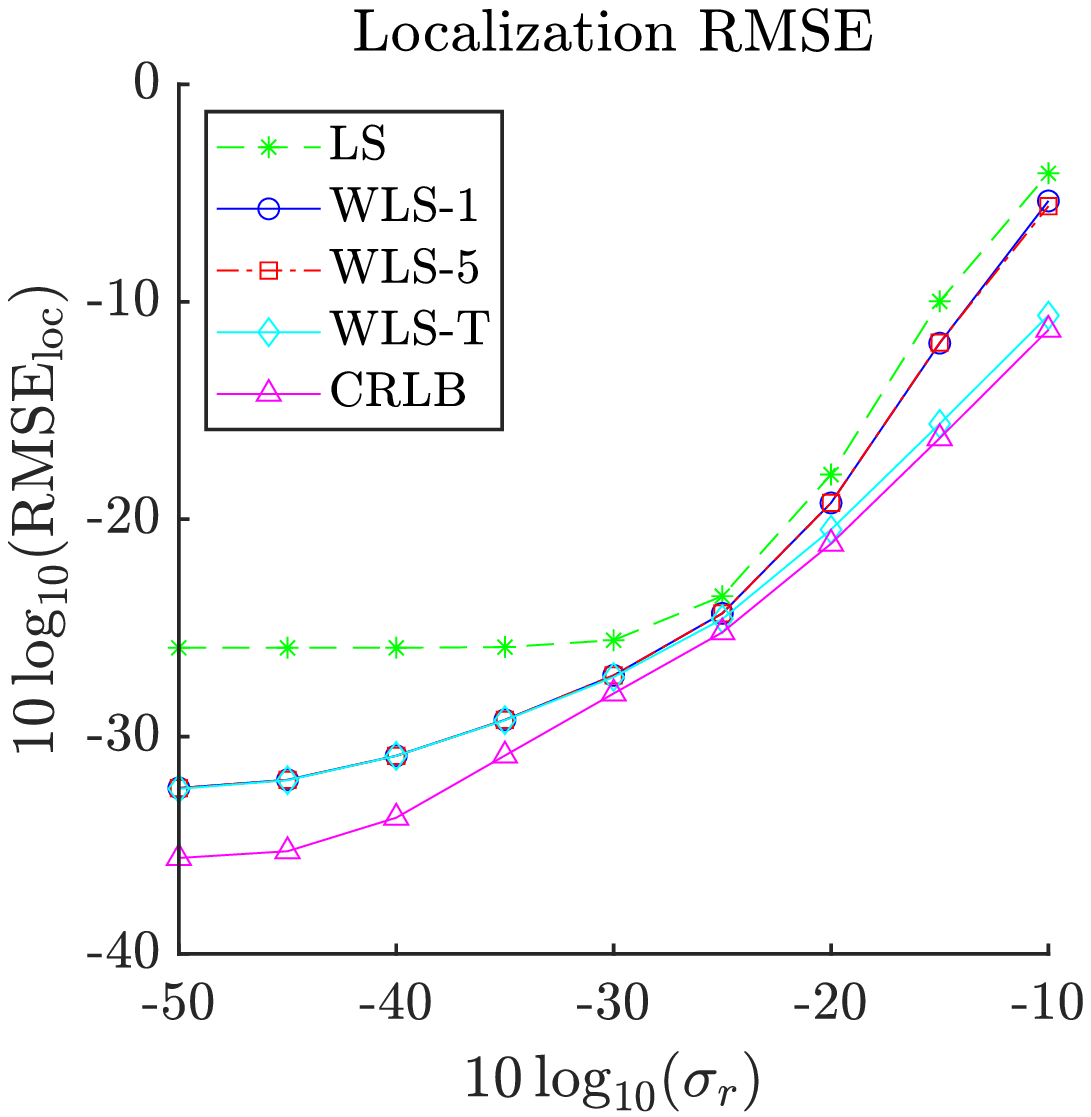}  & \includegraphics[width=0.23\textwidth]{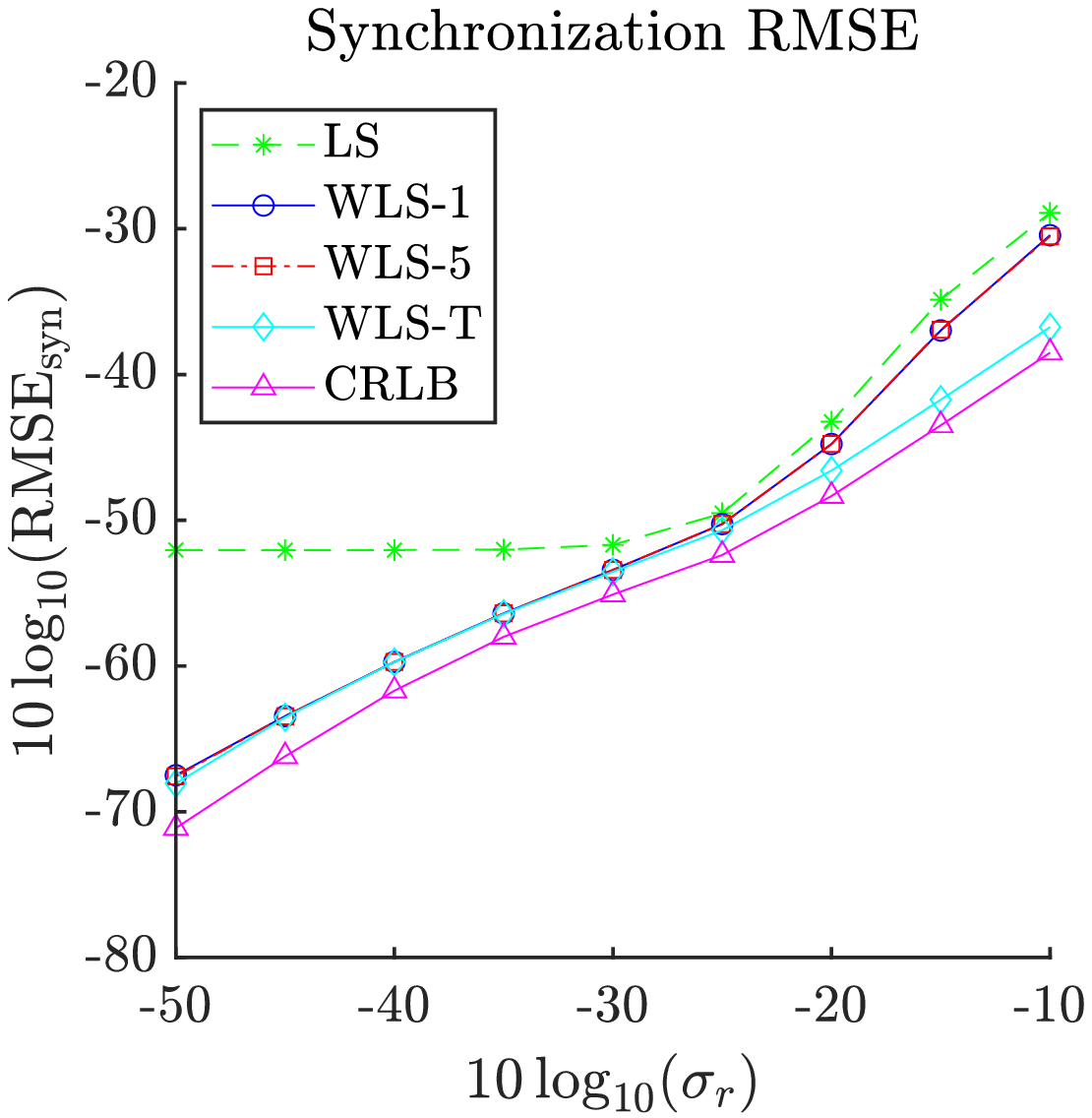}  & \includegraphics[width=0.23\textwidth]{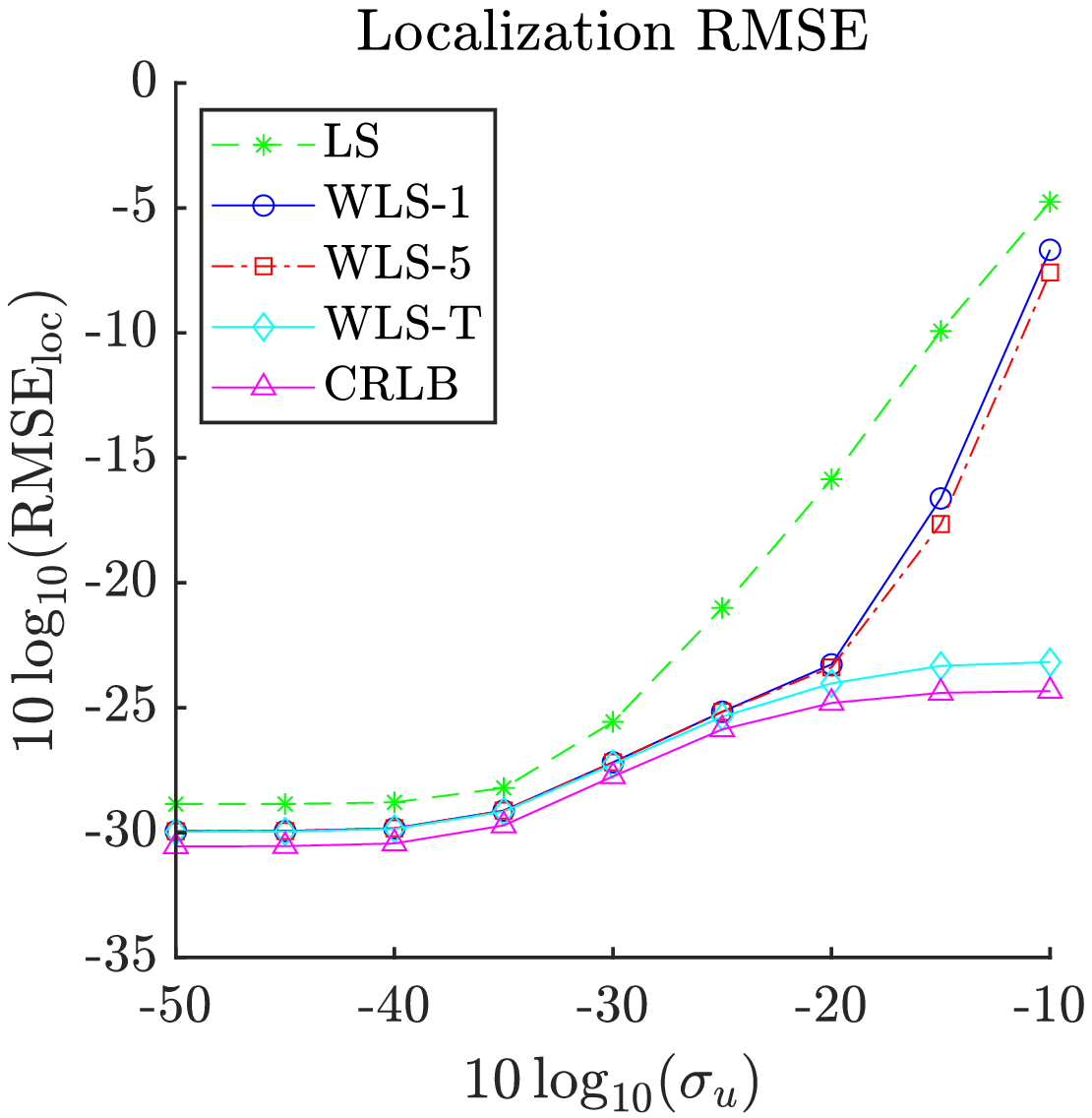}  & \includegraphics[width=0.23\textwidth]{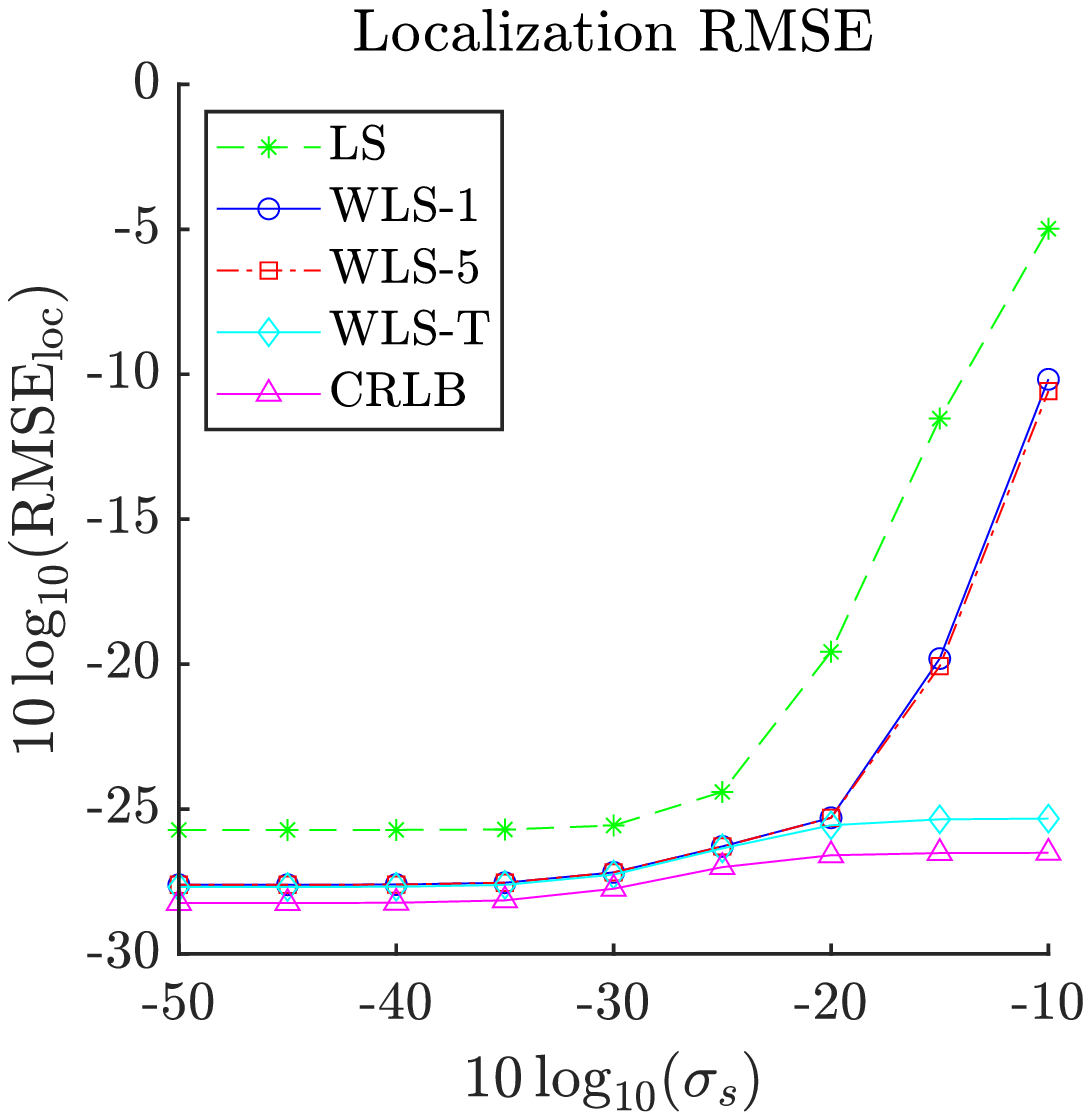}\\
    \small \hspace{0.5cm}(a) & \small \hspace{0.5cm} (b) & \small \hspace{0.5cm} (c) & \small \hspace{0.5cm} (d) \vspace{0.2cm}\\ 
    \includegraphics[width=0.23\textwidth]{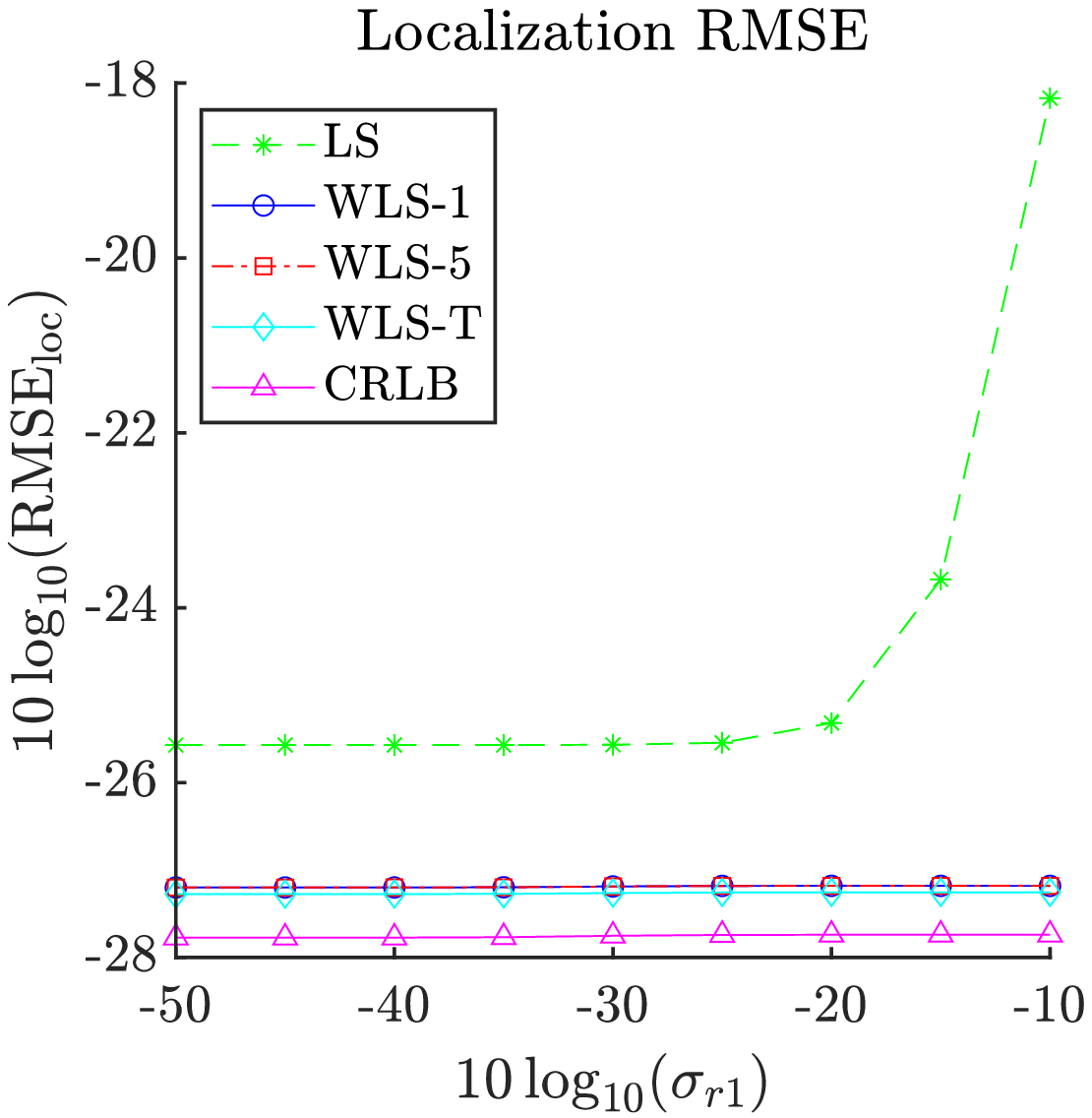}  & 
    \includegraphics[width=0.23\textwidth]{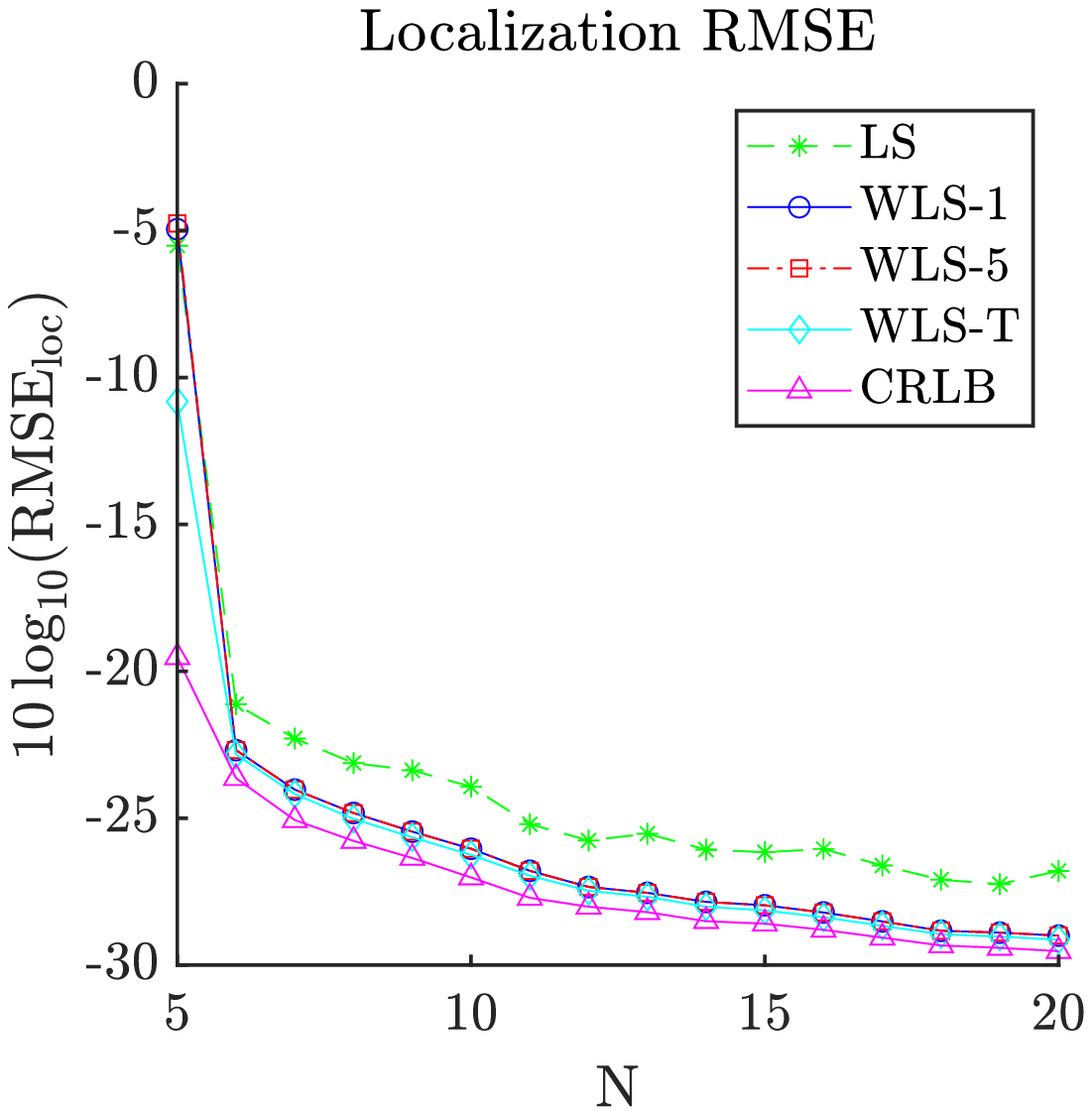}  & 
    \includegraphics[width=0.23\textwidth]{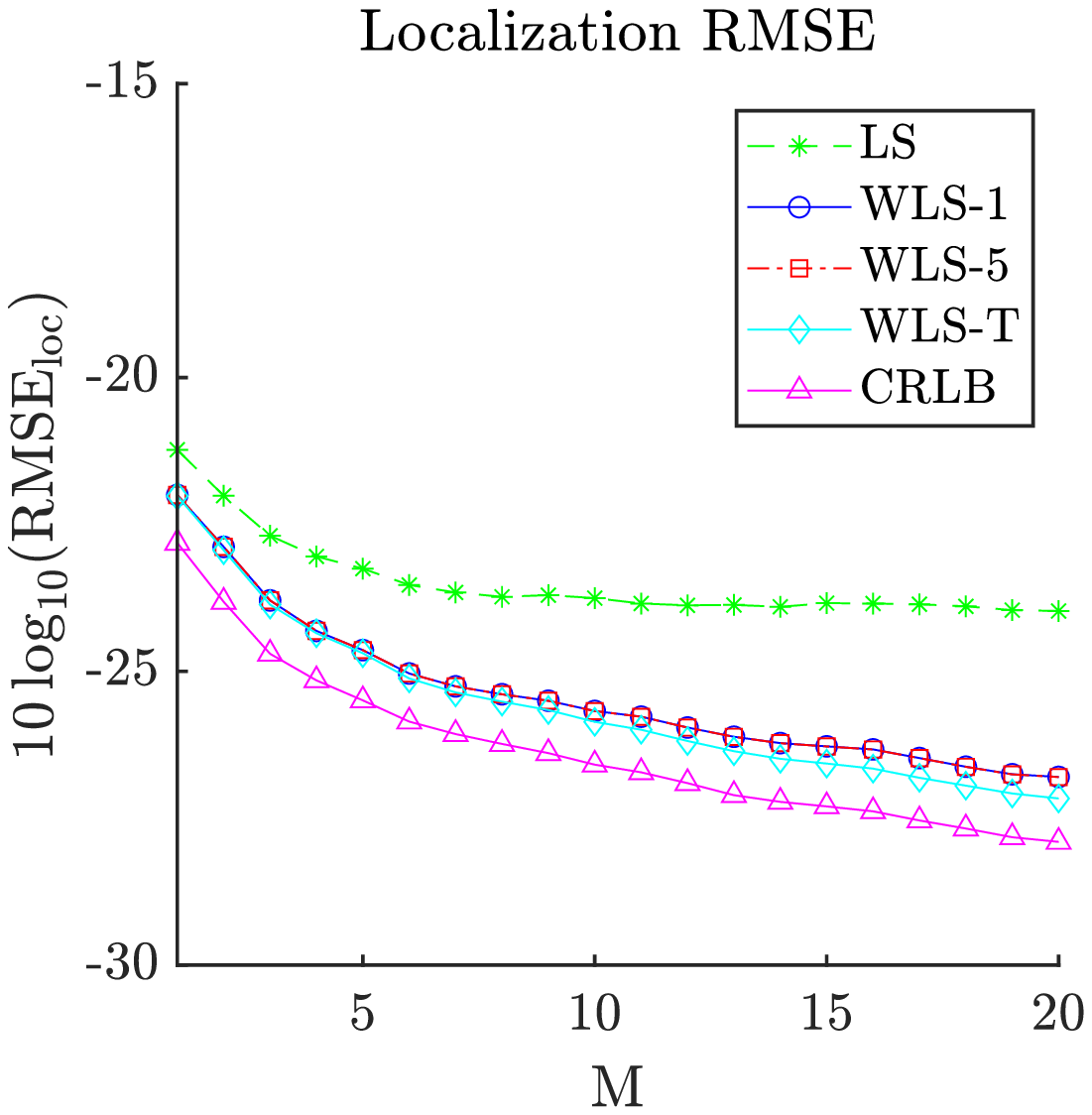}  & 
    \includegraphics[width=0.23\textwidth]{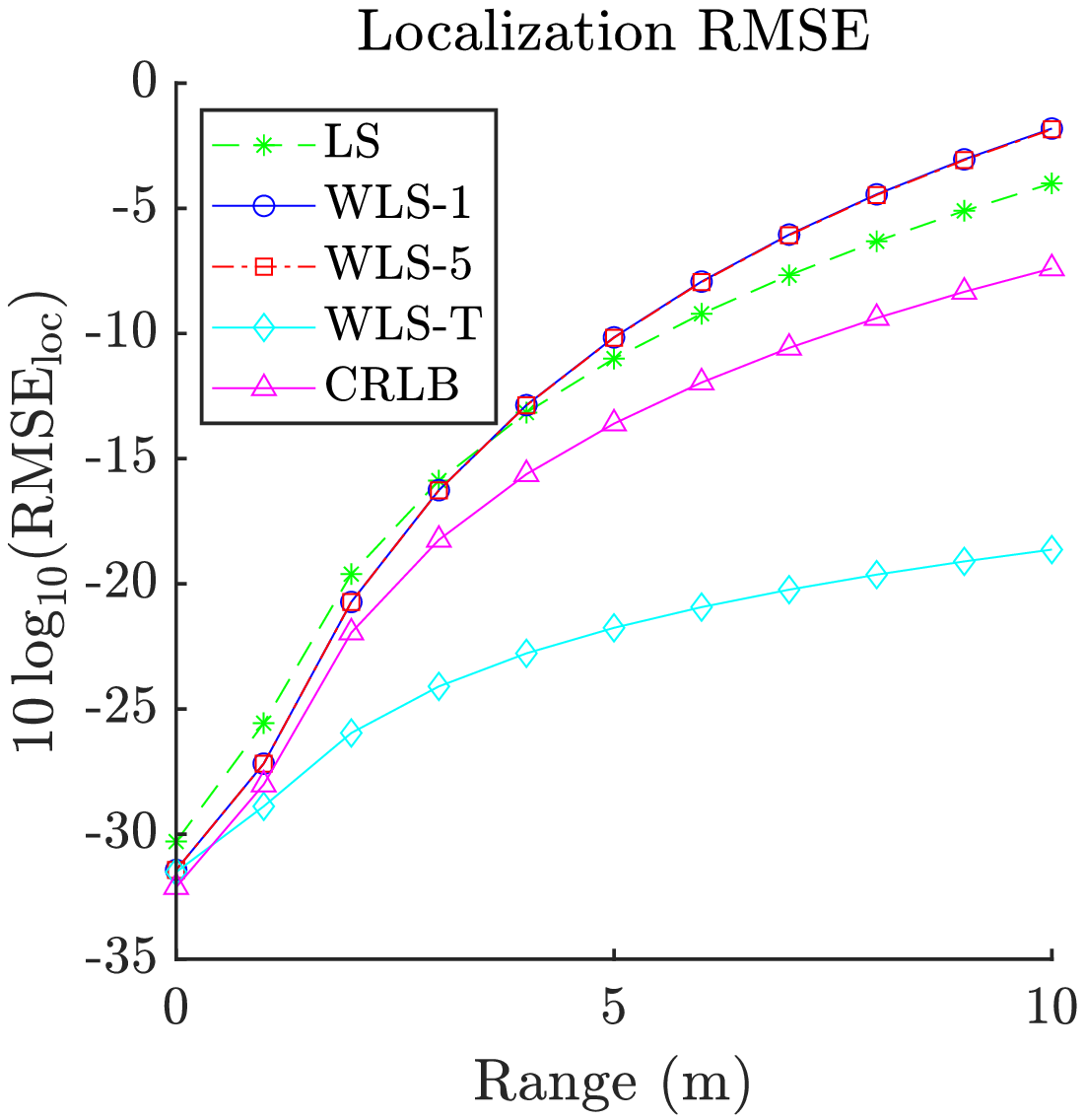}\\
    \small \hspace{0.5cm} (e) & \small \hspace{0.5cm} (f) & \small \hspace{0.5cm} (g) & \small \hspace{0.5cm} (h)
  \end{tabular}
  \caption{Experiments 1-7. Performance of LS, WLS-1, WLS-5, WLS-T, and CRLB under different calibration scenarios. (a) Localization and (b) synchronization RMSEs for varying noise SD in TDOA measurements. (c) Localization RMSE for varying noise SD in emitter 3D position estimates. (d) Localization RMSE for varying noise SD in sensor 3D position estimates. (e) Localization RMSE for varying noise SD in TDOA measurements due to the first sensor only. (f) Localization RMSE for a varying number of emitters, $N$. (g) Localization RMSE for a varying number of sensors, $M$. (h) Localization RMSE for a varying range of the new sensor with respect to the WASN and emitters, $R$.}
  \label{fig:simulations}
\end{figure*}

In the first experiment, we evaluated the performance of each method when varying $10\log_{10}(\sigma_r)$ in $[-50,-10]$ and keeping all other parameters fixed. The results in Fig. \ref{fig:simulations} (a) and (b) show that WLS-1 and WLS-5 achieve equivalent localization and synchronization performance. Additionally, it is shown that WLS-1, WLS-5, WLS-T, and CRLB tend to behave comparably for low-to-moderate noise conditions, and they all significantly outperform LS. However, when the noise increases, the RMSE of the proposed method starts to deviate from its theoretical estimate and the CRLB. Throughout the rest of the experiments, it was found that the relative behavior of different approaches remained similar for both localization and synchronization RMSEs. Henceforth, we will mostly show and discuss the localization RMSE only for conciseness.\par

In the second and third experiments, we individually vary $10\log_{10}(\sigma_u)$ and $10\log_{10}(\sigma_s)$ in $[-50,-10]$, respectively, while keeping all other parameters fixed. The results in Fig. \ref{fig:simulations} (c) and (d) show a somewhat similar pattern to those of the first experiment, with the exception that the gap between the proposed method and its theoretical estimate plus the CRLB widens much more considerably at high noise values. The highly noticeable gap is attributed to the assumption of low noise in (\ref{eq:uj_si}), which was made in the approximation of (\ref{eq:delta_eta_ij}) derived in Appendix \ref{appendix2}. When comparing the results of the first three experiments, it is also evident that the impact of $\sigma_r$ is more significant compared to $\sigma_u$ and $\sigma_s$. Hence, a reliable method for TDOA estimation is crucial to achieving satisfactory calibration performance.\par

Let $\sigma_{r1}$ be the standard deviation of the AWGN in TDOA measurements due to the first sensor only, i.e., $\sigma_{r1}$ is the standard deviation of $\Delta r_{11}, ..., \Delta r_{1N}$ in (\ref{eq:rij}). In the fourth experiment, we let $\mathbf{Q}_r = \text{blkdiag}\left\{\sigma_{r1}^2\mathbf{I}_{N},\sigma_r^2\mathbf{I}_{N(M-1)}\right\}$ and verified the performance of the different methods when varying $10\log_{10}(\sigma_{r1})$ in $[-50,-10]$ while keeping $10\log_{10}(\sigma_{r})$ fixed at $-30$. The main objective of this experiment is to validate the performance of WLS under uneven measurement noise. Fig. \ref{fig:simulations} (e) shows that varying $\sigma_{r1}$ has no effect on the results of WLS and CRLB, which can be attributed to a large number of reliable TDOA measurements and accurate knowledge of noise second-order statistics. The performance of LS, on the other hand, deteriorates noticeably at higher values of $\sigma_{r1}$, as expected.\par

The fifth experiment evaluates each method's performance as the number of emitters, $N$, varies in $[5,20]$. Fig. \ref{fig:simulations} (f) shows that once $N$ exceeds $5$, which is the minimum number of emitters required, WLS outperforms LS and attains comparable performance to the CRLB. Otherwise, when $N = 5$, we see an excessively wide gap between the CRLB and the remaining methods. Similarly, the sixth experiment consists in evaluating the performance of the different methods as the number of sensors, $M$, varies in $[1,20]$. The results in Fig. \ref{fig:simulations} (g) show that reliable localization performance is possible even for $M = 1$. It is also evident, once again, that WLS outperforms LS and attains comparable performance to the CRLB.\par

In the seventh experiment, we evaluate the performance of different methods as the range of a sensor that needs to be calibrated, $R$, varies in $[0,10]$ (m). The objective of this experiment is to observe the effect on calibration performance as the position of a new sensor with respect to the WASN and emitters changes from near-field to far-field. The results in Fig. \ref{fig:simulations} (h) show that as $R$ starts to exceed the aperture of the WASN and emitters, $A$, the performance of all methods deteriorates significantly, and the results of WLS-T become unreliable. The reason for such performance degradation is that calibration performance is highly susceptible to noise in TDOA measurements at far-field scenarios. This is also an issue in TDOA-based source localization, where it is known that reliable localization is not quite possible in far-field scenarios and instead, the DOA is generally of interest instead \cite{tdoa_far}. The reason for the latter can be attributed to the numerous approximations made in deriving (\ref{eq:wls-T}), which fail to consider the effect of far-field scenario.\par

\begin{figure}
  \centering
  \begin{tabular}{ c c }
    \includegraphics[width=0.225\textwidth]{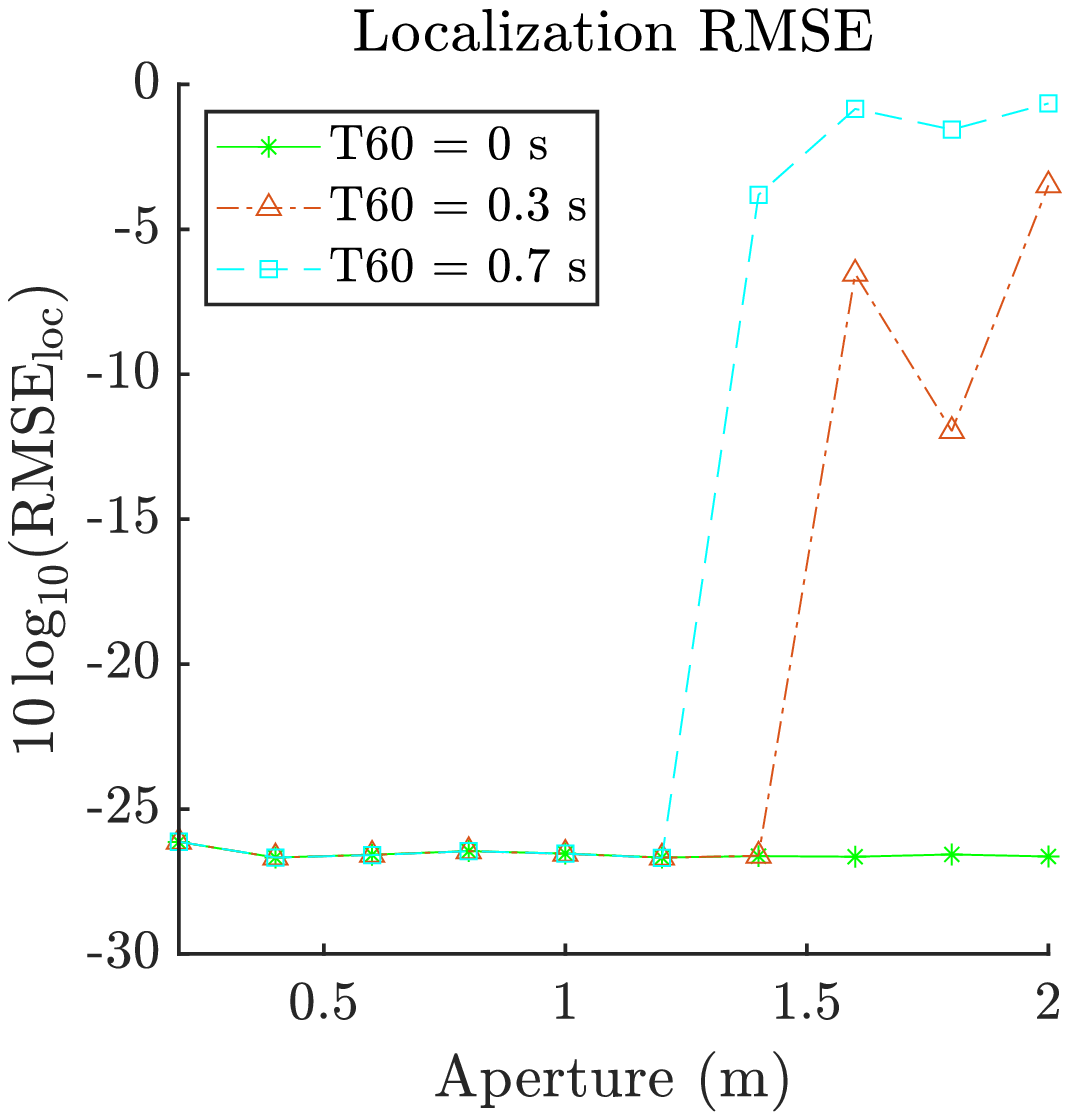}  & \includegraphics[width=0.225\textwidth]{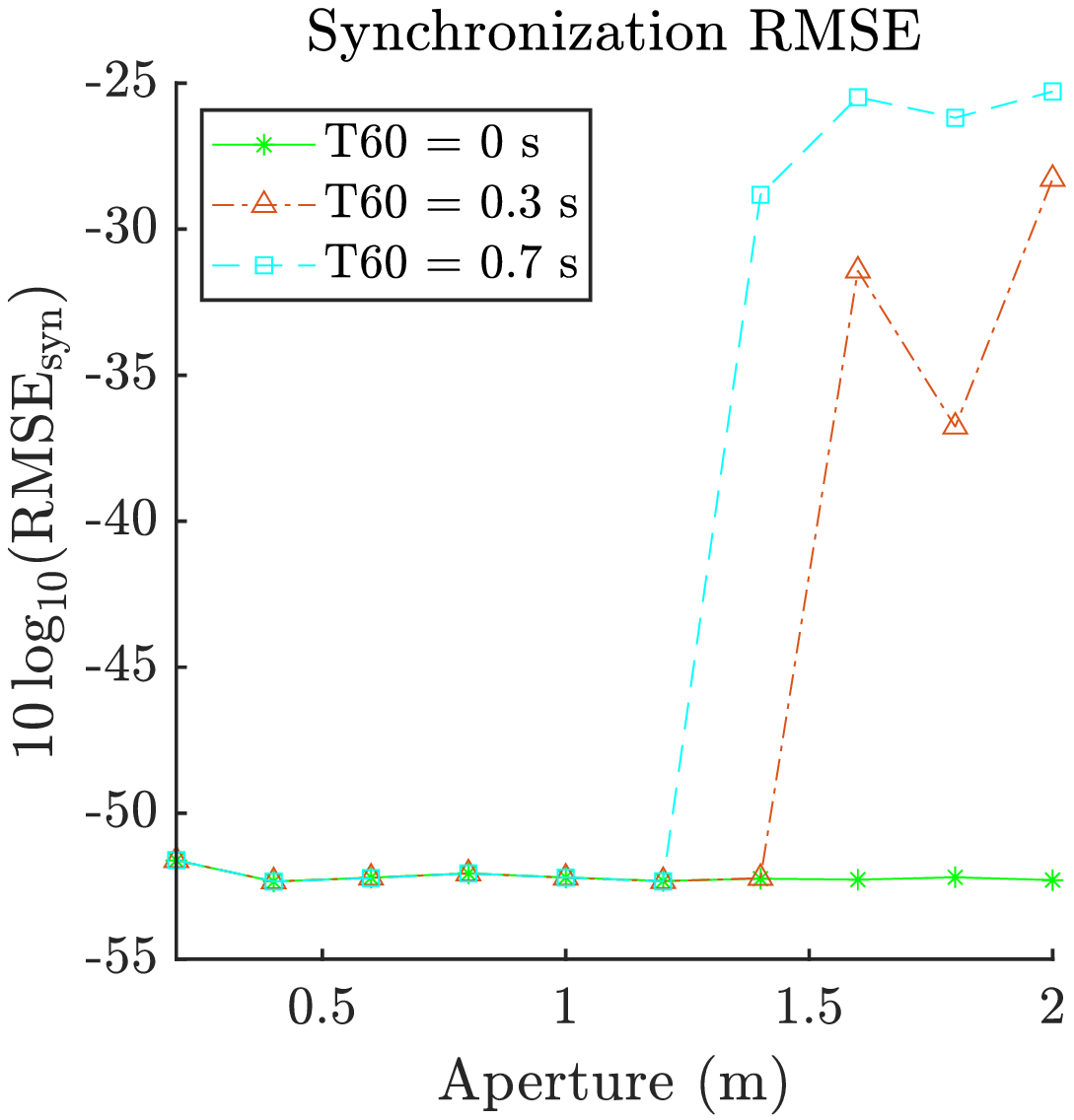} \\
    \small \hspace{0.5cm} (a) & \small \hspace{0.5cm} (b)
  \end{tabular}
  \caption{Experiment 8. Acoustic simulations. (a) Localization and (b) synchronization RMSEs of WLS-1 for varying reverberation times and apertures of the WASN and emitters.}\label{fig:real_simul_dist}
\end{figure}

The eighth and final experiment consists of a series of acoustic simulations. The purpose is to analyze the effect on the calibration performance of WLS-1 for varying reverberation times ($T_{60}$) and apertures of the WASN and emitters. In this experiment, a room of size $5 \times 5 \times 3$ m was simulated using the image method in \cite{habets}. The WASN, emitters and new sensor were all placed randomly in the center of the room using the same procedure as in the previous experiments. All elements were constrained to be at least $0.5$ m away from the walls. Three different reverberation times were considered by letting $T_{60} \in \left\{0,0.3,0.7\right\}$ s. For each value of $T_{60}$, both $A$ and $R$ were set to vary jointly in $[0.2,2]$ m. The calibration signal consisted of white noise with a duration of one second and was sampled at $f_s = 48$ kHz. The signals captured across individual sensors were simulated by convolving the calibration signal with a corresponding room impulse response (RIR) of length equal to $1024$ samples. TDOAs were estimated using GCC-PHAT \cite{GCC} plus quadratic interpolation \cite{tashev2009sound}. These were then corrupted with a given $\tau_p$ to adequately simulate the asynchronous TDOA measurements in (\ref{eq:tau_ij}). The rest of the parameters, $c$, $M$, $N$, $\sigma_s$, $\sigma_u$, $\sigma_r$, $N_s$ and $N_i$, were set to theirs, previously defined values (i.e., as the default values). Due to a lack of literature on the estimation of variance in TDOA measurements computed using GCC-PHAT, for simplicity, we assumed AWGN with SD $\sigma_r = c/f_s \approx 0.71$ cm, which is equivalent to an error of one sample. Consequently, the corresponding noise covariance matrices were defined in the same manner as in the previous experiments.\par

The results of the eighth experiment are plotted in Fig. \ref{fig:real_simul_dist}. It can be observed that when $T_{60} = 0$, the increase in aperture has no negative effect on the calibration performance of WLS-1. The same observation holds for $T_{60} > 0$ up to a threshold where TDOA estimates become unreliable. The mentioned threshold is $A = R = 1.6$ m when $T_{60} = 0.3$ s and $A = R = 1.4$ when $T_{60} = 0.7$ s. Hence, it can be noted that the value of maximum aperture allowing reliable calibration performance decreases for increasing reverberation time. To mitigate this limitation, more sophisticated TDOA estimation methods under reverberant conditions should be used instead.\par

\section{Conclusion} \label{sec:conclusion}
 The availability of a large number of calibrated sensors is generally essential for WASNs to perform optimally. This study proposed a closed-form solution for joint calibration and synchronization of a sensor node, which may not include a built-in acoustic emitter, with respect to a WASN. As such, the proposed method is useful for efficiently scaling a WASN as new sensors become available and for providing a means to calibrate and synchronize sensor nodes lacking built-in emitters. The presented method exploits signals from spatially distributed sources to acquire TDOA measurements between the existing WASN and a new sensor. The problem is modeled as a system of multivariate nonlinear TDOA equations, which is then converted into a system of linear equations through a simple transformation. Next, WLS is applied to estimate the position and synchronization offset of the new sensor with respect to the existing WASN. Simulation results showed the following concerning the performance of the proposed estimator: applying WLS only once (WLS-1) is sufficient to achieve satisfactory results; the estimator is robust to noise in all types of measurements given second-order noise statistics are known a priori; overall performance is often comparable to the CRLB provided measurements noise is not excessive; calibration performance tends to improve for increasing number of WASN sensors and/or emitters; the new sensor should preferably be placed at near-field with respect to the WASN and emitters; the use of a robust TDOA estimation method is crucial under highly-reverberant conditions, especially when the aperture of the sensors and emitters is large; otherwise the combination of GCC-PHAT plus quadratic interpolation is sufficient for highly-accurate calibration performance.\par

% The proposed method can handle calibration of multiple sensors in parallel or sequentially and supports the calibration of sensor nodes without built-in emitters.

%% before appendix (optional) and bibliography:
\section*{Acknowledgment}
This work was supported by the National Institute on Deafness and Other Communication Disorders (NIDCD) of the National Institutes of Health (NIH) under Award 5R01DC015430-05. The content is solely the responsibility of the authors and does not necessarily represent the official views of the NIH. A provisional patent on this work has been obtained with USPTO, Serial number: 63/301,867.

% ---------------------------------------------------------------------
% Appendix  (optional)
\appendices
\section{Linearization of error in $\tilde{\mathbf{\eta}}$}
\label{appendix2}
Let, 
% We are interested in knowing the error in $\eta_{ij}$, if there exist $\boldsymbol\Delta\mathbf{u}_j$ and $\boldsymbol\Delta\mathbf{s}_i$ amount of uncertainties in the position of $\mathbf{u}_j$ and $\mathbf{s}_i$ respectively. More specifically, we want to find approximation of $\Delta \eta_{ij}$, which can be written linearly in terms of  $\boldsymbol\Delta\mathbf{u}_j$ and, $\boldsymbol\Delta\mathbf{s}_i$ . Noisy euclidean distance can be given as 
\begin{equation}
     \tilde{\eta}_{ij} = ||\mathbf{\tilde{u}}_j - \mathbf{\tilde{s}}_i|| = \eta_{ij} + \Delta \eta_{ij} .
\end{equation}
Applying (\ref{eq:si}) and (\ref{eq:uj}) results in the following relationship
% Separating true and noisy terms, we have
\begin{equation} \label{eq:eta_delta_eta}
     \eta_{ij} + \Delta \eta_{ij} = ||(\mathbf{u}_j + \boldsymbol\Delta\mathbf{u}_j) - (\mathbf{s}_i + \boldsymbol\Delta\mathbf{s}_i)|| ,
\end{equation}
where we wish to express $\Delta \eta_{ij}$ as a linear combination of $\boldsymbol\Delta\mathbf{u}_j$ and $\boldsymbol\Delta\mathbf{s}_i$. Squaring both sides in (\ref{eq:eta_delta_eta}) and neglecting the second-order error terms on the right-hand side, we get
% Due to non-linear nature of (\ref{eq:eta_delta_eta}), it is not obvious to separate the error term on the right hand side (RHS). Hence, we take square on both sides of (\ref{eq:eta_delta_eta}) and neglect the second order error term on the RHS to get,
\begin{equation} \label{eq:nij2}
\begin{gathered}
     \Delta \eta_{ij}^2 + 2\eta_{ij} \Delta \eta_{ij} - 2(\mathbf{u}_j - \mathbf{s}_i)^T(\boldsymbol\Delta\mathbf{u}_j - \boldsymbol\Delta\mathbf{s}_i) = 0 . 
\end{gathered}
\end{equation}
Solving the quadratic equation results in 
\begin{equation} \label{eq:quadratic_eta}
    \begin{multlined}
        \Delta \eta_{ij} = -\eta_{ij} \pm \eta_{ij}\sqrt{1 + \frac{2(\mathbf{u}_j - \mathbf{s}_i)^T(\boldsymbol\Delta\mathbf{u}_j - \boldsymbol\Delta\mathbf{s}_i)}{\eta_{ij}^{2}}} .
    \end{multlined}
\end{equation}
Now, assuming $\Delta \eta_{ij} \ll \eta_{ij}$ the solution with a negative sign can be ignored. This assumption also implies that
\begin{equation}
    \dfrac{2(\mathbf{u}_j - \mathbf{s}_i)^T(\boldsymbol\Delta\mathbf{u}_j - \boldsymbol\Delta\mathbf{s}_i)}{\eta_{ij}^{ 2}} \ll 1 .
\end{equation}
Consequently, applying the Maclaurin series expansion to (\ref{eq:quadratic_eta}) up to linear terms only results in the following approximation 
% In (\ref{eq:quadratic_eta}), we can eliminate a solution with $``-"$ sign, by assuming $\Delta \eta_{ij} \ll \eta_{ij} $. This assumption also states that $\frac{2(\mathbf{u}_j - \mathbf{s}_i)^T(\boldsymbol\Delta\mathbf{u}_j - \boldsymbol\Delta\mathbf{s}_i)}{\eta_{ij}^{ 2}} \ll 1$. Hence further applying Maclaurin series to (\ref{eq:quadratic_eta}), we get 
\begin{equation}
     \Delta \eta_{ij} \approx \frac{(\mathbf{u}_j - \mathbf{s}_i)^T(\boldsymbol\Delta\mathbf{u}_j - \boldsymbol\Delta\mathbf{s}_i)}{\eta_{ij}} .
\end{equation}
% here, second and higher order terms in Maclaurin series are neglected and, $\Delta \eta_{ij}$ represents the linear approximation of actual error in $\eta_{ij}$.

\section{Partial derivatives in CRLB}
\label{appendix1}
Since the CRLB matrix is symmetric, it suffices to find solutions to the partial derivatives in either the lower or upper triangular portion of (\ref{eq:crlb}) only. Solutions to the partial derivatives in the lower triangular portion are given by
\begin{equation}
    \begin{aligned}
     \mathbb{E}\left[ 
        \frac{\partial^2 \ln{p(\mathbf{v}|\boldsymbol\theta)}}{\partial\boldsymbol\upgamma \partial\boldsymbol\upgamma^{T}} 
    \right] &= -\left( \frac{\partial \mathbf{r}}{\partial \boldsymbol\upgamma} \right)^T \mathbf{Q}_r^{-1} \left( \frac{\partial \mathbf{r}}{\partial \boldsymbol\upgamma} \right) \\[5pt]
      \mathbb{E}\left[ 
        \frac{\partial^2 \ln{p(\mathbf{v}|\boldsymbol\theta)}}{\partial\boldsymbol\upgamma \partial\mathbf{u}^{ T}} 
    \right] &= -\left( \frac{\partial \mathbf{r}}{\partial \boldsymbol\upgamma} \right)^T \mathbf{Q}_r^{-1} \left( \frac{\partial \mathbf{r}}{\partial \mathbf{u}} \right)\\[5pt]
     \mathbb{E}\left[ 
        \frac{\partial^2 \ln{p(\mathbf{v}|\boldsymbol\theta)}}{\partial\boldsymbol\upgamma \partial\mathbf{s}^T} 
    \right] &= -\left( \frac{\partial \mathbf{r}}{\partial \boldsymbol\upgamma} \right)^T \mathbf{Q}_r^{-1} \left( \frac{\partial \mathbf{r}}{\partial \mathbf{s}} \right)\\[5pt]
     \mathbb{E}\left[ 
        \frac{\partial^2 \ln{p(\mathbf{v}|\boldsymbol\theta)}}{\partial \mathbf{u} \partial\mathbf{s}^T} 
    \right] &= -\left( \frac{\partial \mathbf{r}}{\partial \mathbf{u}} \right)^T \mathbf{Q}_r^{-1} \left( \frac{\partial \mathbf{r}}{\partial \mathbf{s}} \right)\\[5pt]
      \mathbb{E}\left[ 
        \frac{\partial^2 \ln{p(\mathbf{v}|\boldsymbol\theta)}}{\partial \mathbf{u} \partial\mathbf{u}^T} 
    \right] &= -\left( \frac{\partial \mathbf{r}}{\partial \mathbf{u}} \right)^T \mathbf{Q}_r^{-1} \left( \frac{\partial \mathbf{r}}{\partial \mathbf{u}} \right) - \mathbf{Q}_u^{-1}\\
    \mathbb{E}\left[ 
        \frac{\partial^2 \ln{p(\mathbf{v}|\boldsymbol\theta)}}{\partial \mathbf{s} \partial\mathbf{s}^T} 
    \right] &= -\left( \frac{\partial \mathbf{r}}{\partial \mathbf{s}} \right)^T \mathbf{Q}_r^{-1} \left( \frac{\partial \mathbf{r}}{\partial \mathbf{s}} \right) - \mathbf{Q}_s^{-1},
\end{aligned}
\end{equation}
where
\begin{equation}
    \begin{aligned}
        \frac{\partial \mathbf{r}}{\partial\boldsymbol\upgamma} &=
        \begin{bmatrix}
            \mathbf{A}_1^T &
            \cdots &
            \mathbf{A}_M^T
        \end{bmatrix}^T \\[5pt]
        \mathbf{A}_i &=
        \begin{bmatrix}
            \frac{(\mathbf{u}_1 - \mathbf{p})^T}{||\mathbf{u}_1 - \mathbf{p}||} & 1 \\
            \vdots                           & \vdots \\
            \frac{(\mathbf{u}_N - \mathbf{p})^T}{||\mathbf{u}_N - \mathbf{p}||} & 1
        \end{bmatrix},
    \end{aligned}
    \label{eq:r_partial_gamma}
\end{equation}

\begin{equation}
\begin{aligned}
    \frac{\partial \mathbf{r}}{\partial\mathbf{u}} &=
    \begin{bmatrix}
        \mathbf{A}_1^T &
        \cdots &
        \mathbf{A}_M^T
    \end{bmatrix}^T  \\[5pt] \mathbf{A}_i &=
    \begin{bmatrix}
       \mathbf{a}_{i1}^T &  &  \\
        & \ddots & \\
         &  & \mathbf{a}_{iN}^T
    \end{bmatrix} \\[5pt]
     \mathbf{a}_{ij} &= \frac{(\mathbf{u}_j - \mathbf{s}_i)}{||\mathbf{u}_j - \mathbf{s}_i||} - \frac{(\mathbf{u}_j - \mathbf{p})}{||\mathbf{u}_j - \mathbf{p}||},
    \label{eq:r_partial_u}
    \end{aligned}
\end{equation}
\begin{small}
\begin{equation}
    \begin{aligned}
        \frac{\partial \mathbf{r}}{\partial\mathbf{s}} &=
        \begin{bmatrix}
            \mathbf{A}_1^T &
            \cdots &
            \mathbf{A}_M^T
        \end{bmatrix}^T \\[5pt]
        \mathbf{A}_i &=
        \begin{bmatrix}
            \mathbf{0}_{1 \times 3(i-1)} & -\frac{(\mathbf{u}_1 - \mathbf{s}_i)^T}{||\mathbf{u}_1 - \mathbf{s}_i||} & \mathbf{0}_{1 \times 3(M-i)} \\
            \vdots & \vdots & \vdots \\
            \mathbf{0}_{1 \times 3(i-1)} & -\frac{(\mathbf{u}_N - \mathbf{s}_i)^T}{||\mathbf{u}_N - \mathbf{s}_i||} & \mathbf{0}_{1 \times 3(M-i)}
        \end{bmatrix}.
    \end{aligned}
    \label{eq:r_partial_s}
\end{equation}
\end{small}

\ifCLASSOPTIONcaptionsoff
  \newpage
\fi

\bibliographystyle{IEEEtran}

\bibliography{new} 

%=======================================================

\end{document}